\documentclass[aps,prb,showpacs,twocolumn]{revtex4}
\usepackage{bm,color,amsmath,amssymb,mathrsfs,latexsym,graphicx,psfrag}
\usepackage{pst-all,epsfig,wrapfig,subfigure,rotating,wasysym} 

\newcommand{\bs}[1]{\boldsymbol{#1}}
\newcommand{\bp}{\begin{pmatrix}}
\newcommand{\ep}{\end{pmatrix}}

\newcommand{\half}{$\frac{1}{2}$ }
\newcommand{\bR}{\bs{R}}
\newcommand{\bk}{\bs{k}}

\newcommand{\ket}[1]{\left|#1\right\rangle}
\newcommand{\bra}[1]{\left\langle #1 \right|}

\newcommand{\ev}[1]{\bigl\langle#1\bigr\rangle}

\newcommand{\sket}[1]{|#1\rangle}
\newcommand{\sbra}[1]{\langle #1 |}

\newcommand{\up}{\uparrow}
\newcommand{\dw}{\downarrow}

\def\bd{\begin{displaymath}}
\def\ed{\end{displaymath}}
\def\be{\begin{equation}}
\def\ee{\end{equation}}
\def\bea{\begin{eqnarray}}
\def\eea{\end{eqnarray}}
\def\bi{\begin{itemize}}
\def\ei{\end{itemize}}
\def\bn{\begin{enumerate}}
\def\en{\end{enumerate}}

\def\ie{{\it i.e.},\ }
\def\eg{{\it e.g.}\ }
\def\ea{{\it et al.}}
\def\etc{\emph{etc.}\ }

\allowdisplaybreaks[1]

\begin{document}
\title{Magnetic Excitations in the Site-Centered Stripe Phase: \\
  Spin Wave Theory of Coupled Three-Leg Ladders} \author{Martin
  Greiter and Holger Schmidt} \affiliation{Institut f\"ur Theorie der
  Kondensierten Materie and DFG Center for Functional Nanostructures (CFN),
  KIT, Campus S\"ud, D 76128 Karlsruhe} \pagestyle{plain}
\date{\today}

\begin{abstract}
%
  The success of models of coupled two-leg spin ladders in describing
  the magnetic excitation spectrum of La$_{2-x}$Ba$_x$CuO$_4$ had been
  interpreted previously as evidence for bond-centered stripes.  In a
  recent article, however, we have determined the magnetic coupling
  induced by the charge stripes between bond- or site-centered spin
  stripes modeled by two- or three-leg ladders, respectively.  We
  found that only the site-centered models order.  We further
  indicated excellent agreement of a fully consistent analysis of
  coupled three-leg ladders using a spin wave theory of bond with the
  experimental data.  Here, we provide a full and detailed account of
  this analysis.
\end{abstract}

%
%
%
%

\maketitle

\section{Introduction}

Twenty years after the discovery, the mechanism of high-$T_{\text{c}}$
superconductivity in the copper oxide materials is still considered
one of the most important outstanding problem in contemporary
physics~\cite{zaanen-06np138,orenstein-00s468}.  The materials are
described by mobile charge carriers (holes) doped into a
quasi-twodimensional spin 1/2 antiferromagnet~\cite{Zhang-88prb3759,
  Eskes-88prl1415}.  Inelastic neutron scattering experiments have
revealed a magnetic resonance
peak~\cite{fong-95prl316,bourges-00s1234} and, in some compounds,
periodic modulations in the spin and charge density
(stripes)~\cite{zaanen-89prb7391,kato-90jpsj1047,tranquada-95n561,emery-99pnas8814,zaanen-01pmb1485,mook-02prl097004,kivelson-03rmp1201,berg-09njp115004}.
Tranquada \ea~\cite{tranquada-04n534} found that the magnetic
excitation spectrum of stripe ordered La$_{1.875}$Ba$_{0.125}$CuO$_4$
looks similar to disordered
YBa$_2$Cu$_3$O$_{6+x}$~\cite{bourges-00s1234} or
Bi$_2$Sr$_2$CaCu$_2$O$_{8+\delta}$~\cite{fauque-07prb214512}, and
observed that the data are consistent with bond-centered stripes
modeled by two-leg ladders.  This experiment is considered of key
importance for the field, as it it may provide the decisive hint as to
within which framework copper oxide superconductors may be understood.

With regard to such a framework, there is no consensus at present, but
a fierce competition among different schools of thought.  One of these
schools~\cite{emery-99pnas8814,zaanen-01pmb1485,kivelson-03rmp1201,berg-09njp115004}
attributes the unusual properties of the doped, two-dimensional
antiferromagnets to their propensity to form stripes, or their
proximity to a quantum critical point (QCP) at which stripe order sets
in.  The resulting picture is highly appealing.  Static stripes have
been observed~\cite{tranquada-95n561,mook-02prl097004} only in certain
compounds, most notably La$_{2-x}$Sr$_{x}$CuO$_4$ at a hole doping
concentration $x=\frac{1}{8}$, and are known to suppress
superconductivity.  On the other hand, the mere existence of stripes
would impose an effective one-dimensionality, and hence provide a
framework to formulate fractionally quantized excitations.  This
one-dimensionality would be roughly consistent with an enormous body
of experimental data on the cuprates, including the electron spectral
functions seen in angle-resolved photo emission spectroscopy (ARPES).
The charge carriers, the holons, would predominantly reside in the
charge stripes, as they could maximize their kinetic energy in these
antiferromagnetically disordered regions.  In the spin stripes, by
contrast, the antiferromagnetic exchange energy between the spins
would be maximized, at the price of infringing on the mobility of the
charge carriers.  Most importantly, the spin stripes would impose a
coupling between the charge stripes, which would yield an effective,
pairwise confinement between the low-energy spinon and holon
excitations residing predominantly in the charge stripes.  The
mechanism of confinement would be similar to that of coupled spin
chains or spin
ladders~\cite{dagotto-96s618,shelton-96prb8521,greiter02prb054505}.
The holes would be described by spinon-holon bound states, and the
dominant contribution to the magnetic response measured in Tranquada's
as well as all other neutron scattering experiments would come from
spinon-spinon bound states.

\begin{figure}[t]
  \begin{center}
    \includegraphics[height=65mm]{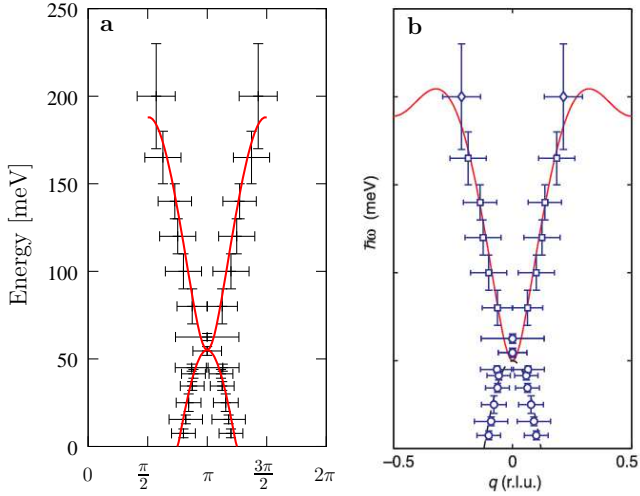}
  \end{center}
  \caption{(Color online) (a) Superpositions of cuts along $(k_x,\pi)$
    and $(\pi,k_y)$ for the lowest mode $\omega (\bk)$ obtained
    with the bond operator spin wave theory of coupled three-leg 
    ladders presented here 
    (red) superimposed with the experimental data obtained by
    inelastic neutron scattering by Tranquada
    \ea~\cite{tranquada-04n534} (black).  (b) The neutron data as
    originally presented, with a triplon dispersion of a two-leg
    ladder superimposed (red line) (Reprinted by permission from
    Macmillian Publishers Ltd: Nature 429: 534-538, \copyright 2005).}
  \label{fig:tddtr}
\end{figure}

\begin{figure}[t]
  \begin{center}
    \includegraphics[width=0.56\linewidth]{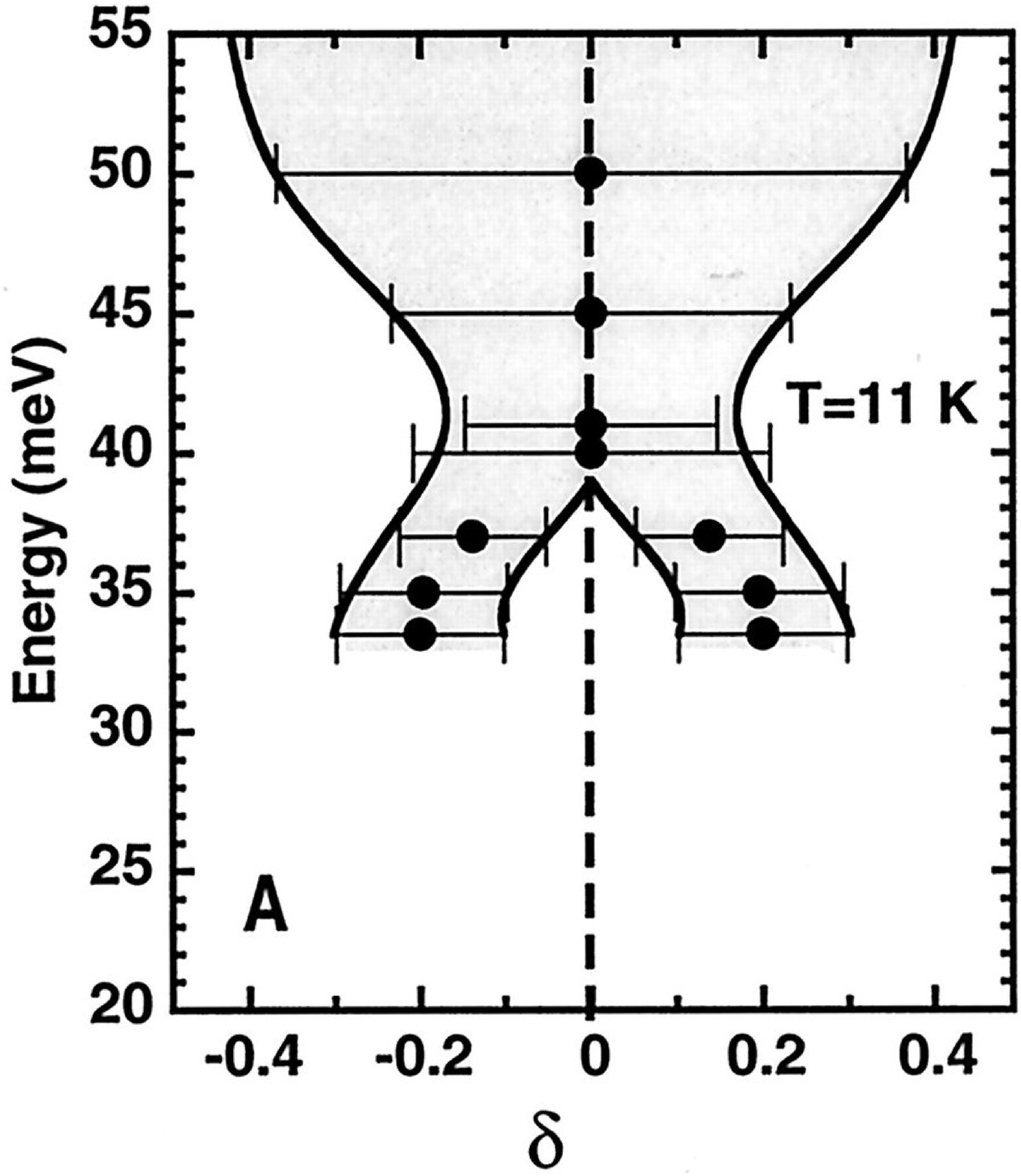}
  \end{center}
  \vspace{-16pt}
  \caption{Overall momentum dependence of the magnetic response of
    superconducting YBa$_2$Cu$_3$O$_{6.85}$ as reported by Bourges
    \ea\cite{bourges-00s1234}.  There is no stripe ordering in this
    compound (From Bourges \ea, Science 288: 1234-1237 (2000).
    Reprinted by permission from AAAS).}
  \label{fig:elephant}
\end{figure}

The similarity of the ``hour-glass'' spectrum shown in Fig.~4b of
Tranquada \ea~\cite{tranquada-04n534} (which is reproduced for
comparison in Fig.~\ref{fig:tddtr}b) with the ``elephants trousers''
observed by Bourges et al.~\cite{bourges-00s1234} (which are
reproduced for comparison in Fig.~\ref{fig:tddtr}b) provides the most
striking evidence in favor of the picture advocated by this school,
which attributes the anomalous properties of generic, disordered CuO
superconductors to the formation of dynamic (rather than static)
stripes, which fluctuate on time scales which are slow compared to the
energy scales of most experimental probes.  This picture is
considered to receive additional support by Xu
\ea~\cite{Xu-07prb014508}, who observed that the magnetic response of
La$_{1.875}$Ba$_{0.125}$CuO$_4$ at higher energies is independent of
temperature, while the stripe order melts at about $T_{\rm
  st}\sim54{\rm K}$.  Measurements on `untwinned' samples of
YBa$_2$Cu$_3$O$_{6.6}$, where one would expect the dynamical stripes
to orient themselves along one of the axis, however, exhibit a strong
anisotropy in the response only at energies below the
resonance~\cite{hinkov-04n650}, while the response is fourfold
rotationally symmetric at higher energies~\cite{hinkov-07np780}.  We
believe, however, that this only indicates that the formation of
stripe correlations, be it static or dynamic, is a low energy
phenomenon, while the high energy response probes itinerant
antiferromagnets at length scales on which the stripes are essentially
invisible.

\begin{figure}[h]
  \begin{minipage}[c]{0.45\textwidth}
    \includegraphics[width=\linewidth]{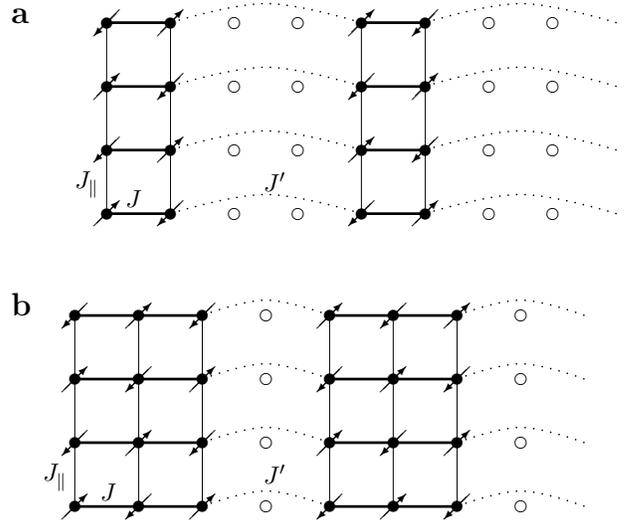}
  \end{minipage}
  \caption{Spin model for (a) bond-centered and (b) site-centered
    stripes in CuO superconductors.}
  \label{fig:skizze}
\end{figure}

An extremely appealing feature of the experiment by Tranquada
\ea~\cite{tranquada-04n534} is that it immediately suggests a model of
ferromagnetically coupled two-leg ladders, as the upper part of the
measured spectrum agrees strikingly well with the triplon (or
spinon-spinon bound state) mode of isolated two-leg ladders (see
Fig.~\ref{fig:tddtr}b).  The experiment hence appeared to point to
bond-centered rather than site-centered stripes, (\ie stripes as
depicted in Fig.~\ref{fig:skizze}a rather than
Fig.~\ref{fig:skizze}b), and thereby to resolve a long outstanding
issue.

This interpretation received support by theoretical
studies~\cite{vojta-04prl127002,uhrig-04prl267003,anisimov-04prb172501}.
Vojta and Ulbricht~\cite{vojta-04prl127002} used a bond operator
formalism~\cite{sachdev-90prb9323} similar to ours to study a
spin-only model of stripes~\cite{krueger-03prb134512} of coupled
two-leg ladders (as depicted in Fig.~\ref{fig:skizze}a) with
$J_{\parallel}=J$, took into account a bond-boson
renormalization~\cite{eder98prb12832} of $J$, and assumed a value $J'$
for the ferromagnetic coupling between the ladders which is large
enough to close the spin gap of the ladders, \ie to induce long range
magnetic order.  Within their approximations, a value of $J'=-0.06J$
is sufficient.  This value is not consistent with previous
studies~\cite{gopalan-94prb8901,tworzydlo-99prb115,dalosto-00prb928},
but as no method to calculate or even estimate the true $J'$ induced
by charge stripes had been available, it did not seem a problem at the
time.  The spectrum they obtained agrees well with experimental data
measured by Tranquada \ea~\cite{tranquada-04n534}, and hence appeared
to justify their assumptions \emph{a posteriori}.  They concluded in
favor of bond-centered stripes.
This conclusion was independently strengthened by Uhrig, Schmidt, and
Gr\"uninger~\cite{uhrig-04prl267003}, who used the method of
continuous unitary transformations to study a model of
ferromagnetically coupled two-leg ladders, and observed that the
critical value of $J_{\rm c}'$ can be significantly reduced if a
cyclic exchange term $J_{\rm cyc}$ on the ladders is
included~\cite{nunner-02prb180404}.  They likewise fine-tuned $J'$ to
the QCP where the gap closes and long-range magnetic order ensues, and
reported good agreement with experiment.

On the other hand, Seibold and
Lorenzana~\cite{seibold-05prl107006,seibold-06prb144515} calculated
the magnetic response for a range of dopings within the time-dependent
Gutzwiller approximation, and found good agreement with the measured
data for both bond- and site-centered stripe models.

In a recent article~\cite{greiter-10prb144509}, we investigated
whether it is reasonable to assume that the ferromagnetic coupling
$J'$ induced by the charge stripe between the spin stripes modelled by
two-leg ladders is sufficiently to induce long range order.  There are
several estimates for the critical value $J_{\rm c}'$ required if the
coupling between isotropic ladders is antiferromagnetic in the
literature.  Gopalan, Rice, and Sigrist~\cite{gopalan-94prb8901} find
$J_{\rm c}'\approx 0.25J$ in a simple mean-field treatment of
bond-bosons.  Quantum Monte Carlo (QMC) calculations by Tworzyd\l{}o
\ea~\cite{tworzydlo-99prb115} yielded $J_{\rm c}'=0.30(2)J$, a value
subsequently confirmed by Dalosto and Riera\cite{dalosto-00prb928}.
We redid the mean-field calculation of Gopalan
\ea~\cite{gopalan-94prb8901} for ferromagnetic (FM) couplings $J_{\rm
  c}'<0$, and found that within this approximation, the absolute value
of $J_{\rm c}'$ is independent of the sign of the coupling.  QMC
calculations by Dalosto \ea~\cite{dalosto-00prb928}, however, indicate
that the true value is at least $J_{\rm c}=-0.4J$ (see Fig.\ 6b of
their article).  The physical reason why a significant coupling
between the ladders is required to induce magnetic order is that the
individual two-leg ladders possess a gap of order $\Delta\approx J/2$.
As a cyclic exchange term $J_{\rm cyc}\approx 0.25J$ reduces this gap
by a factor of two~\cite{nunner-02prb180404}, we expected that $J_{\rm
  c}'$ would likewise be reduced by a factor of two.  We hence
concluded that a FM coupling of at least somewhere between $J_{\rm
  c}'=-0.2J$ and $-0.4J$ is required, depending on the strength of a
possible cyclic exchange term.

The value we obtained for the ferromagnetic coupling induced by the
charge stripes between the spin stripes through exact diagonalization
of small clusters with and without charge stripes, however, is
$J'=-0.05J$~\cite{greiter-10prb144509}.  The details of this
calculation are given in Appendix \ref{sec:jprime}.  The coupling is
hence insufficient to induce order in a model of coupled two-leg
ladders describing bond-centered stripes.  This does not imply that
the stripes cannot be bond-centered, but rather implies that it is not
sensible to describe bond-centered stripes through spin-only models
of coupled two-leg ladders.

For a model of site-centered stripes described by
antiferromagnetically coupled three-leg ladders, as shown in
Fig.~\ref{fig:skizze}b, the critical coupling required for long range
order to set in is by contrast $J_{\rm c}'=0$.  The reason is simply
that there is no need to close a gap, as the three-leg ladders are
individually gapless~\cite{dagotto-96s618}.  A conventional spin wave
analysis for such a spin-only model of three- and four-leg ladders was
performed by Yao, Carlson, and Campbell~\cite{yao-06prl017003}, who
found that their approximation agrees reasonable well with the
experimental data if they take $J'=0.05J$ and $J'=-0.09J$ for coupled
three- and four-leg ladders, respectively.  The calculation we present
in Appendix \ref{sec:jprime}, however, singles out $J'=0.07J$ for the
antiferromagnetic coupling between spin stripes modeled by three-leg
ladders.  In our previous work~\cite{greiter-10prb144509}, we
announced that a fully consistent spin wave theory of bond operators
representing the eight-dimensional Hilbert spaces on each rung of the
three-leg ladders agrees perfectly with the experimental data if and
only if the correct, calculated value $J'=0.07J$ is used for the
coupling.

In this context, one may ask whether it might be possible to obtain an
equally valid description in terms of bond-centered stripes modelled by
four-leg ladders.  We believe the answer is no, as the width of the
charge stripes in between the ladders would be zero, and one
would have to assume that the antiferromagnetic coupling $J$ between
neighboring sites of the original $t$-$J$ model, would turn into a
weak ferromagnetic coupling between the four-leg ladders.  There would
be no foundation for such an assumption.  Furthermore, since the
four-leg ladders are gapped, just as the two-leg ladders are, one
would need to fine-tune this ferromagnetic coupling to exactly
the point where the gap closes and magnetic order with $k=\pi\pm\pi/4$
emerges.  So regardless of the agreement with the measured spectrum
one might be able to obtain, we believe that a spin-only model of
four-leg ladders would not constitute a valid theory.

In this paper, we provide a full and detailed account of our analysis
of our spin-only model of coupled three-leg ladders.  The paper is
organized as follows.  In section \ref{sec:basis}, we introduce a
basis as well as a set of bosonic creation and annihilation operators
for the three-site rungs of the ladders, in terms of which we write
both the rung Hamiltonian and the spin operators on the individual
sites.  In section \ref{sec:coupling}, we couple the rungs both along
the ladders and across neighboring ladders, and self-consistently
determine the fiducial state such that all the terms linear in a
single creation or annihilation operator in the resulting Hamiltonian
vanish.  In section \ref{sec:expanding}, we rewrite this Hamiltonian
in terms of momentum space operators, and expand it to bi-linear order
in term of those.  We then solve for the low energy spectrum using a
multi-dimensional Bogoliubov transformation in section
\ref{sec:solution}.  In section \ref{sec:discussion}, we compare our
results to the experimental data obtained by Tranquada
\ea~\cite{tranquada-04n534}, and investigate the dependence of the
spectrum we obtain on the value of the inter-ladder coupling $J'$.
Finally, we present our conclusions in section \ref{sec:conclusions}.

\section{Basis states for three-site rungs}
\label{sec:basis}

\begin{figure}[h]
  \begin{minipage}[c]{0.45\textwidth}
    \includegraphics[width=0.27\linewidth]{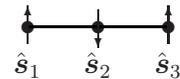}
  \end{minipage}
  \caption{Single rung on sublattice $\cal A$}
  \label{fig:single}
\end{figure}

To begin with, consider a single rung of a three-leg ladder,
consisting of spins which are antiferromagnetically coupled with a
coupling $J$ we set to unity (see Fig.\ \ref{fig:single}).  
For later purposes, let us consider a rung belonging to sublattice
$\cal A$, \ie set up conventions the rungs belonging to sublattice
$\cal A$ will inherit in the following sections.  Denoting the SU(2)
spin operators for the three spin $\frac{1}{2}$'s on the sites by
$\bs{s}_1$,$\bs{s}_2$, and $\bs{s}_3$, the Hamiltonian for the rung
reads
\begin{equation}
\hat H^{\cal A} = \hat{\bs{s}}_1\hat{\bs{s}}_2 + \hat{\bs{s}}_2\hat{\bs{s}}_3 . 
\end{equation}
Diagonalization yields the following eigenvalues and eigenvectors:
\begin{eqnarray}\nonumber
&E=-1& 
\begin{cases}
  |b_{-1/2}\rangle \hspace{-5pt}&\hspace{-2pt}=
  -\frac{1}{\sqrt{6}}\bigl(|\!\up\dw\dw\rangle
  -2|\!\dw\up\dw\rangle
  +|\!\dw\dw\up\rangle)\\
  |b_{1/2}\rangle \hspace{-5pt}&\hspace{-2pt}=
  -\frac{1}{\sqrt{6}}\bigl(|\!\dw\up \up \rangle
  -2|\!\up  \dw \up  \rangle
  +|\!\up \up \dw \rangle\bigr)
\end{cases}\\[3pt]
&E=0&
\begin{cases}
  |a_{-1/2}\rangle \hspace{-5pt}&\hspace{-2pt}=
  \frac{1}{\sqrt{2}}\bigl(|\!\up\dw\dw\rangle
    - |\!\dw\dw\up\rangle\bigr)\\
  |a_{1/2}\rangle \hspace{-5pt}&\hspace{-2pt}=
  \frac{1}{\sqrt{2}}\bigl(|\!\dw\up\up\rangle
    -|\!\up \up \dw \rangle\bigr)
\end{cases}\\[3pt] \nonumber
&E=\frac{1}{2}&
\begin{cases} 
  |c_{-3/2}\rangle \hspace{-5pt}&\hspace{-2pt}=
    |\!\dw\dw\dw\rangle\\
  |c_{-1/2}\rangle \hspace{-5pt}&\hspace{-2pt}=
  \frac{1}{\sqrt{3}}\bigl(|\!\dw\dw\up\rangle
    + |\!\dw \up \dw  \rangle
    + |\!\up \dw \dw \rangle\bigr)\\
  |c_{1/2}\rangle \hspace{-5pt}&\hspace{-2pt}=  
  \frac{1}{\sqrt{3}}\bigl(|\!\up \up \dw  \rangle
    + |\!\up \dw \up  \rangle
    + |\!\dw \up \up  \rangle\bigr)\\
  |c_{3/2}\rangle  \hspace{-5pt}&\hspace{-2pt}=
    |\!\up \up \up \rangle \, .
\end{cases}
\end{eqnarray}
Note that the two states $\sket{a_{-1/2}}$ and $\sket{a_{-1/2}}$ are
antisymmetric under spacial reflections interchanging sites 1 and 3 on
the rung, while all other states are symmetric.  This distinction will
prove useful when expanding the Hamiltonian for the coupled ladders
in Sec.~\ref{sec:expanding} below.

We denote the orthonormal basis formed by these eight states by
\begin{eqnarray}\label{eq:m}
M &=&  
\{|b_{-1/2}\rangle, |b_{1/2}\rangle, |a_{-1/2}\rangle, |a_{1/2}\rangle,
\nonumber\\ &&\hspace{50pt}
|c_{-3/2}\rangle , |c_{-1/2}\rangle , |c_{1/2}\rangle , |c_{3/2}\rangle \}.
\hspace{10pt}
\end{eqnarray}
In this basis, the Hamiltonian matrix is trivially given by
\begin{eqnarray}\label{eq:hadiadics}\nonumber
\hat H^{\cal A} &=& -\Big(\sket{b_{-1/2}}\sbra{b_{-1/2}}
+\sket{b_{1/2}}\sbra{b_{1/2}}\Big)\\ \nonumber
&+& \frac{1}{2}\Big(\sket{c_{-3/2}}\sbra{c_{-3/2}} 
+\sket{c_{-1/2}}\sbra{c_{-1/2}}\\
&&\hspace{20pt}+\ \sket{c_{-1/2}}\sbra{c_{-1/2}}
+\sket{c_{3/2}}\sbra{c_{3/2}}\Big). 
\end{eqnarray}

Neither of these exact eigenstates, however, is suited as a fiducial
state for spin wave theory.  We are hence led to define a vacuum
state
\begin{eqnarray}\label{eq:fidstate}\nonumber
\sket{\tilde{b}_{-1/2}}
&\equiv& \sket{b_{-1/2}}\,\cos\phi+\sket{c_{-1/2}}\,\sin\phi
    \\[5pt]\nonumber
&=& \textstyle
    \bigl(\sket{\!\up\dw\dw}+\sket{\!\dw\dw\up}\bigr)
    \bigl(-\frac{1}{\sqrt{6}}\cos\phi+\frac{1}{\sqrt{3}}\sin\phi\bigr)
    \\[2pt] 
&&+ \ \textstyle\sket{\!\dw\up\dw}
    \bigl(\sqrt{\frac{2}{3}}\cos\phi+\frac{1}{\sqrt{3}}\sin\phi\bigr),
\end{eqnarray}
which interpolates between the quantum ground state $\sket{b_{-1/2}}$
of the isolated rung with $S^{z}=-\frac{1}{2}$ for $\phi=0$ and the
classically N\'eel ordered state $|\!\!\!\dw\up\dw\rangle$ for
$\phi=\arctan(\frac{1}{\sqrt{2}})=0.6155$.  The  parameter
$\phi$ will depend on the coupling between the rungs and the ladders.
The motivation for introducing the state $\sket{\tilde b_{-1/2}}$ will
become clear as we determine $\phi$ self-consistently below.

\begin{figure}[t]
  \begin{minipage}[c]{0.42\textwidth}
    \includegraphics[width=\linewidth]{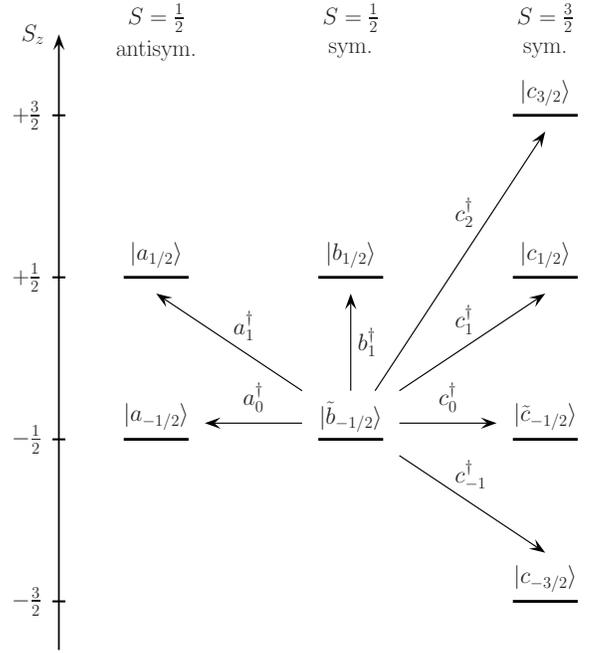}
  \end{minipage}
  \caption{Bosonic operator for sublattice ${\mathcal A}$.}
  \label{fig:operators2}
\end{figure}
Since we wish $\sket{\tilde b_{-1/2}}$ to be one of our basis states, 
we replace \eqref{eq:m} by
\begin{eqnarray}\label{eq:ma}
M^{\cal A} &=& \{\ket{\mu}^{\cal A};\mu=1,\ldots,8\} \nonumber\\[3pt]\nonumber 
&=&
\{|\tilde{b}_{-1/2}\rangle, |b_{1/2}\rangle, |a_{-1/2}\rangle, |a_{1/2}\rangle,
\\  &&\hspace{30pt}
|c_{-3/2}\rangle, |\tilde{c}_{-1/2}\rangle, |c_{1/2}\rangle, |c_{3/2}\rangle\},
\hspace{10pt}
\end{eqnarray}
with 
\begin{eqnarray}\label{eq:rota}
\begin{pmatrix}\sket{b_{-1/2}}\\[3pt]\sket{c_{-1/2}}\end{pmatrix}
&=&\begin{pmatrix}u & -v \\[3pt] v & u \end{pmatrix}
\begin{pmatrix}\sket{\tilde{b}_{-1/2}}\\[3pt]
               \sket{\tilde{c}_{- 1/2}}\end{pmatrix}
\end{eqnarray}
and $u = \cos\phi$, $v = \sin\phi$.
The relevant terms in the Hamiltonian \eqref{eq:hadiadics} transform 
into
\begin{eqnarray}
&\begin{pmatrix}\sket{b_{-1/2}}, & \sket{c_{-1/2}}\end{pmatrix}
\begin{pmatrix}-1 & 0 \\[3pt] 0 & \frac{1}{2}\end{pmatrix}
\begin{pmatrix}\sbra{b_{-1/2}}\\[3pt] \sbra{c_{-1/2}}\end{pmatrix}
\hspace{67pt}&\\[5pt]\nonumber
&
=\begin{pmatrix}\sket{\tilde{b}_{-1/2}}, & \sket{\tilde{c}_{-1/2}}\end{pmatrix}
\begin{pmatrix}\frac{1}{2}-\frac{3}{2}u^2 & \frac{3}{2}uv \\[3pt] 
               \frac{3}{2}uv & -1 + \frac{3}{2}u^2\end{pmatrix}
\begin{pmatrix}\sbra{\tilde{b}_{-1/2}}\\[3pt]\sbra{\tilde{c}_{-1/2}}\end{pmatrix}
.&
\end{eqnarray}

As a next step, we introduce bosonic creation and annihilation
operators $a_0^{\dagger}\equiv\sket{a_{-1/2}}\sbra{\tilde{b}_{-1/2}}$
\etc, as indicated in Fig.~\ref{fig:operators2}.  The subscripts of
these operators refer to the change in the $z$-component of the total
spin on the rung.  Note that these operators do not obey the
commutation relations of independent ladder operators, as we can
create only one ``particle'' with either $a_0^{\dagger}$ or
$a_1^{\dagger}$ or any other creation operator from the ``vacuum''
state $\sket{\tilde{b}_{-1/2}}$.  

Completeness and orthonormality of the basis \eqref{eq:ma} implies
\begin{multline}\label{eq:complete}
  \sket{\tilde{b}_{-1/2}}\sbra{\tilde{b}_{-1/2}} =
  \big(1-b_1^{\dagger}b_1^{\phantom{\dagger}}
  -a_0^{\dagger}a_0^{\phantom{\dagger}}
  -a_1^{\dagger}a_1^{\phantom{\dagger}}\\-c_{-1}^{\dagger}c_{-1}^{\phantom{\dagger}}
  - c_0^{\dagger}c_0^{\phantom{\dagger}} -
  c_1^{\dagger}c_1^{\phantom{\dagger}} -
  c_2^{\dagger}c_2^{\phantom{\dagger}}\big).
\end{multline}
With \eqref{eq:rota} and \eqref{eq:complete}, the rung Hamiltonian
\eqref{eq:hadiadics} may be rewritten in terms of the bosonic operators:
\begin{eqnarray}\nonumber
\hat H^{\cal A}\label{eq:ha} 
&=& \left(\frac{1}{2}-\frac{3}{2}u^2\right) 
+ \left(-\frac{1}{2}+\frac{3}{2}u^2\right)
  \left(a_0^{\dagger}a_0^{\phantom{\dagger}}
    + a_1^{\dagger}a_1^{\phantom{\dagger}}\right)\\\nonumber
&+& \frac{3}{2}uv\left(c_0^{\dagger}+c_0^{\phantom{\dagger}}\right)
+ \frac{3}{2}\left(u^2-v^2\right)c_0^{\dagger}c_0^{\phantom{\dagger}}\\
&-& \frac{3}{2}v^2 b_1^{\dagger}b_1^{\phantom{\dagger}} 
+ \frac{3}{2}u^2\left(c_{-1}^{\dagger}c_{-1}^{\phantom{\dagger}}
+ c_1^{\dagger}c_1^{\phantom{\dagger}} + c_2^{\dagger}c_2^{\phantom{\dagger}}\right).
\end{eqnarray}

\begin{figure}[t]
  \begin{minipage}[c]{0.42\textwidth}
    \includegraphics[width=\linewidth]{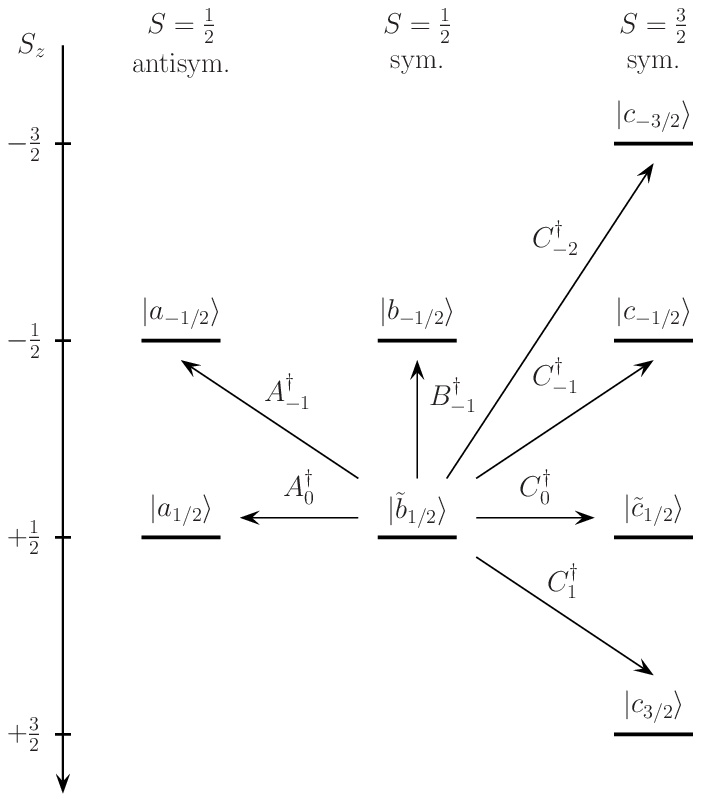}
  \end{minipage}
  \caption{Bosonic operator for sublattice ${\mathcal B}$.}
 \label{fig:operators3}
\end{figure}
On sublattice $\cal B$, we introduce a similar basis $M^{\cal B}$,
with the only difference that the fiducial state
$\sket{\tilde{b}_{1/2}}$ has $S^z=\frac{1}{2}$ instead of
$s^z=-\frac{1}{2}$ for $\sket{\tilde{b}_{-1/2}}$ on sublattice $\cal A$:
\begin{eqnarray}\label{eq:mb}
M^{\cal B} &=& \{\ket{\mu}^{\cal B};\mu=1,\ldots,8\} \nonumber \\[3pt]\nonumber 
&=& 
\{|\tilde{b}_{1/2}\rangle,|b_{-1/2}\rangle, |a_{1/2}\rangle, |a_{-1/2}\rangle,
\\ &&\hspace{30pt}
|c_{3/2}\rangle, |\tilde{c}_{1/2}\rangle, |c_{-1/2}\rangle, |c_{-3/2}\rangle\},
\hspace{10pt}
\end{eqnarray}
with
\begin{eqnarray}\label{eq:rotb}
\begin{pmatrix}\sket{b_{1/2}}\\[3pt]\sket{c_{1/2}}\end{pmatrix} &=& 
\begin{pmatrix}u & -v \\[3pt] v & u \end{pmatrix}
\begin{pmatrix}\sket{\tilde b_{1/2}}\\[3pt]\sket{\tilde c_{ 1/2}}\end{pmatrix}.
\end{eqnarray}
We introduce a second set of bosonic creation and annihilation
operators $A_0^{\dagger}\equiv\sket{a_{1/2}}\sbra{\tilde{b}_{1/2}}$ \etc, as
indicated in Fig.~\ref{fig:operators3}.

The Hamiltonian $H^{\cal B}$ for a single rung belonging to sublattice
$\cal B$ is in analogy to \eqref{eq:ha} given by
\begin{eqnarray}\label{eq;hb}\nonumber
\hat H^{\cal B} &=& 
\left(\frac{1}{2}-\frac{3}{2}u^2\right) 
+ \left(-\frac{1}{2}+\frac{3}{2}u^2\right)
\left(A_{-1}^{\dagger}A_{-1}^{\phantom{\dagger}}
+ A_0^{\dagger}A_0^{\phantom{\dagger}}\right)\\\nonumber
&+&\frac{3}{2}uv\left(C_0^{\dagger}+C_0^{\phantom{\dagger}}\right)
+\frac{3}{2}\left(u^2-v^2\right)C_0^{\dagger}C_0^{\phantom{\dagger}}\\\nonumber
&-&\frac{3}{2}v^2 B_{-1}^{\dagger}B_{-1}^{\phantom{\dagger}}\\
&+&
\frac{3}{2}u^2\left(C_{-2}^{\dagger}C_{-2}^{\phantom{\dagger}}
+C_{-1}^{\dagger}C_{-1}^{\phantom{\dagger}}
+C_{1}^{\dagger}C_{1}^{\phantom{\dagger}}\right).
\end{eqnarray}

For later purposes, we write the spin operators 
$\hat{s}_{\alpha}^{\pm}=\hat{s}_{\alpha}^x\pm i\hat{s}_{\alpha}^y$
and $\hat{s}_{\alpha}^z$ for the individual sites $\alpha=1,2,3$ on
rungs belonging to sublattice $\cal A$ in terms of our bosonic
creation and annihilation operators:
\begin{eqnarray}\label{eq:sa}\nonumber
\hat{s}_{\alpha}^{+}  &=& 
s_{\alpha,21}^{+}b_1^{\dagger}
+s_{\alpha,41}^{+}a_1^{\dagger} + 
s_{\alpha,71}^{+}c_1^{\dagger} + 
s^+_{\alpha,15}c_{-1}^{\phantom{\dagger}}\\\nonumber
&+&
s_{\alpha,82}^{+}c_2^{\dagger}b_1^{\phantom{\dagger}}  
+s_{\alpha,23}^{+}b_1^{\dagger}a_0^{\phantom{\dagger}}
 +s_{\alpha,43}^{+}a_1^{\dagger}a_0^{\phantom{\dagger}} 
\\\nonumber &+&
s_{\alpha,73}^{+}c_1^{\dagger}a_0^{\phantom{\dagger}}
+s_{\alpha,84}^{+}c_2^{\dagger}a_1^{\phantom{\dagger}} 
+s_{\alpha,35}^{+}a_0^{\dagger}c_{-1}^{\phantom{\dagger}} 
\\\nonumber
&+&s_{\alpha,65}^{+}c_0^{\dagger}c_{-1}^{\phantom{\dagger}} 
+s_{\alpha,26}^{+}b_1^{\dagger}c_0^{\phantom{\dagger}} + 
s_{\alpha,46}^{+}a_1^{\dagger}c_0^{\phantom{\dagger}}\\\nonumber
&+&s_{\alpha,76}^{+}c_1^{\dagger}c_0^{\phantom{\dagger}} + 
s_{\alpha,87}^{+}c_2^{\dagger}c_1^{\phantom{\dagger}},
\\
\hat{s}_{\alpha}^{-} &=&  \bigl(\hat{s}_{\alpha}^{+}\bigr)^{\dagger},\\\nonumber
\hat{s}_{\alpha}^z  &=& s^z_{\alpha,11}\big(1-b_1^{\dagger}b_1^{\phantom{\dagger}}
-a_0^{\dagger}a_0^{\phantom{\dagger}}-a_1^{\dagger}a_1^{\phantom{\dagger}}
\\\nonumber 
&&\hspace{30pt}
-c_0^{\dagger}c_0^{\phantom{\dagger}}-c_1^{\dagger}c_1^{\phantom{\dagger}}
-c_{-1}^{\dagger}c_{-1}^{\phantom{\dagger}}-c_2^{\dagger}c_2^{\phantom{\dagger}}\big)
\\\nonumber
&+& s_{\alpha,22}^zb_1^{\dagger}b_1^{\phantom{\dagger}} + 
s_{\alpha,55}^zc_{-1}^{\dagger}c_{-1}^{\phantom{\dagger}} 
+ s_{\alpha,66}^{z}c_0^{\dagger}c_0^{\phantom{\dagger}} \\\nonumber
&+&s_{\alpha,77}^zc_1^{\dagger}c_1^{\phantom{\dagger}} + 
s_{\alpha,88}^zc_2^{\dagger}c_2^{\phantom{\dagger}}\\\nonumber
&+& s^z_{\alpha,13}\bigl(a_0^{\dagger}+a_0^{\phantom{\dagger}}\bigr) 
+ s^z_{\alpha,16}\bigl(c_0^{\dagger}+c_0\bigr)\\\nonumber
&+& s^z_{\alpha,27}\bigl(b_1^{\dagger}c_1^{\phantom{\dagger}}
+c_1^{\dagger}b_1^{\phantom{\dagger}}\bigr) 
+ s^z_{\alpha,24}\bigl(b_1^{\dagger}a_1^{\phantom{\dagger}} + 
a_1^{\dagger}b_1^{\phantom{\dagger}}\bigr)\\\nonumber
&+& s^z_{\alpha,36}\bigl(a_0^{\dagger}c_0^{\phantom{\dagger}} + 
c_0^{\dagger}a_0^{\phantom{\dagger}}\bigr)
+ s^z_{\alpha,47}\bigl(a_1^{\dagger}c_1^{\phantom{\dagger}} + 
c_1^{\dagger}a_1^{\phantom{\dagger}}\bigr).
\end{eqnarray}
The matrix elements 
\begin{equation}
  s_{\alpha,\mu\nu}^{\tau}=\sbra{\mu}\hat{s}_{\alpha}^{\tau}\ket{\nu}^{\cal A}
  \quad\text{with}\quad\tau=+,-,z, 
  \label{eq:matel}
\end{equation}
and $\ket{\mu}^{\cal A}$ as defined in \eqref{eq:ma} are written out
explicitly in Appendix \ref{sec:spmz}.

Similarly, the individual spin-operators $\hat{S}_{\alpha}^{\pm}$ 
on rungs belonging to sublattice $\cal B$ are given by:
\begin{eqnarray}\label{eq:sb}\nonumber ^{\phantom{\dagger}}
\hat{S}_{\alpha}^{-}  &=& 
S_{\alpha,21}^{-}B_{-1}^{\dagger}
+S_{\alpha,41}^{-}A_{-1}^{\dagger} + 
S_{\alpha,71}^{-}C_{-1}^{\dagger} + 
S^-_{\alpha,15}C_{1}^{\phantom{\dagger}}\\\nonumber
&+&
S_{\alpha,82}^{-}C_{-2}^{\dagger}B_{-1}^{\phantom{\dagger}}  
+S_{\alpha,23}^{-}B_{-1}^{\dagger}A_0^{\phantom{\dagger}}
 +S_{\alpha,43}^{-}A_{-1}^{\dagger}A_0^{\phantom{\dagger}} 
\\\nonumber &+&
S_{\alpha,73}^{-}C_{-1}^{\dagger}A_0^{\phantom{\dagger}}
+S_{\alpha,84}^{-}C_{-2}^{\dagger}A_{-1}^{\phantom{\dagger}} 
+S_{\alpha,35}^{-}A_0^{\dagger}C_{1}^{\phantom{\dagger}} 
\\\nonumber
&+&S_{\alpha,65}^{-}C_0^{\dagger}C_{1}^{\phantom{\dagger}} 
+S_{\alpha,26}^{-}B_{-1}^{\dagger}C_0^{\phantom{\dagger}} + 
S_{\alpha,46}^{-}A_{-1}^{\dagger}C_0^{\phantom{\dagger}}\\\nonumber
&+&S_{\alpha,76}^{-}C_{-1}^{\dagger}C_0^{\phantom{\dagger}} + 
S_{\alpha,87}^{-}C_{-2}^{\dagger}C_{-1}^{\phantom{\dagger}},
\\
\hat{S}_{\alpha}^{+} & = & \bigl(\hat{S}_{\alpha}^{-}\bigr)^{\dagger},\\\nonumber
\hat{S}_{\alpha}^z   & = & S^z_{\alpha,11}\big(1-
B_{-1}^{\dagger}B_{-1}^{\phantom{\dagger}}-
A_0^{\dagger}A_0^{\phantom{\dagger}}-A_{-1}^{\dagger}A_{-1}^{\phantom{\dagger}}\\\nonumber &&\hspace{30pt}
-C_0^{\dagger}C_0^{\phantom{\dagger}}-C_{-1}^{\dagger}C_{-1}^{\phantom{\dagger}}-
C_1^{\dagger}C_1^{\phantom{\dagger}}-C_{-2}^{\dagger}C_{-2}^{\phantom{\dagger}}
\big)\\\nonumber
&+& S_{\alpha,22}^zB_{-1}^{\dagger}B_{-1^{\phantom{\dagger}}} 
+ S_{\alpha,55}^zC_1^{\dagger}C_1^{\phantom{\dagger}} + 
S_{\alpha,66}^{z}C_0^{\dagger}C_0^{\phantom{\dagger}}\\\nonumber 
&+&S_{\alpha,77}^zC_{-1}^{\dagger}C_{-1}^{\phantom{\dagger}} +
S_{\alpha,88}^zC_{-2}^{\dagger}C_{-2}^{\phantom{\dagger}}\\\nonumber
&+& S^z_{\alpha,13}\bigl(A_0^{\dagger}+A_0^{\phantom{\dagger}}\bigr) + 
S^z_{\alpha,16}\bigl(C_0^{\dagger}+C_0^{\phantom{\dagger}}\bigr)\\\nonumber
&+& S^z_{\alpha,27}\bigl(B_{-1}^{\dagger}C_{-1}^{\phantom{\dagger}}+C_{-1}^{\dagger}B_{-1}^{\phantom{\dagger}}\bigr)\\\nonumber 
&+& S^z_{\alpha,24}\bigl(B_{-1}^{\dagger}A_{-1}^{\phantom{\dagger}}+A_{-1}^{\dagger}B_{-1}^{\phantom{\dagger}}\bigr)\\\nonumber
&+& S^z_{\alpha,36}\bigl(A_0^{\dagger}C_0^{\phantom{\dagger}}+C_0^{\dagger}A_0^{\phantom{\dagger}}\bigr)\\\nonumber
&+& S^z_{\alpha,47}\bigl(A_{-1}^{\dagger}C_{-1}^{\phantom{\dagger}}+C_{-1}^{\dagger}A_{-1}^{\phantom{\dagger}}\bigr),
\end{eqnarray}
with $S_{\alpha,\mu\nu}^{\tau}= 
\sbra{\mu}\hat{S}_{\alpha}^{\tau} \ket{\nu}^{\cal B}$ likewise given in
Appendix \ref{sec:spmz}.

\section{Coupling the rungs}
\label{sec:coupling}

\begin{figure}[t]
  \vspace{5pt}
  \begin{minipage}[c]{0.45\textwidth}
    \includegraphics[width=\linewidth]{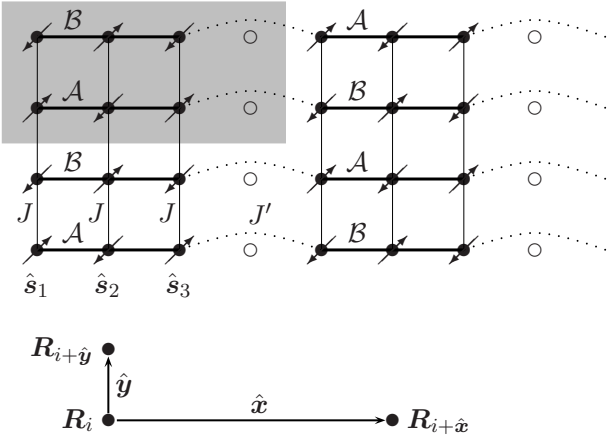}
  \end{minipage}
  \caption{The microscopic model for site centered spin stripes, with
    intra-rung couplings set to unity, inter-rung-intra-ladder
    couplings $J$, and inter-ladder couplings $J'$.  The real space
    unit cell contains two rungs and is indicated by the shaded area
    in gray.}
  \label{fig:2dm}  
\end{figure}

As a microscopic model for site centered spin stripes, we couple the
three-site rungs into three-leg ladders, with the spins coupled
antiferromagnetically with $J$ along the ladders and with $J'$ between
neighboring ladders, as shown in Fig.~\ref{fig:2dm}.  
The sublattice indices assigned to each rung alternate in both
directions, \ie under translation by either of the primitive lattice
vectors $\hat{\bs{x}}=(4a,0)$ or $\hat{\bs{y}}=(0,a)$, where $a$ is the
lattice constant we herewith set to unity.  The microscopic model is
hence given by the Hamiltonian
\begin{eqnarray}\label{eq:h2d}\nonumber
\hat H &=& \sum_{i\in {\cal A}}\left(\hat H^{\cal A}_i 
+ J\sum_{\alpha=1}^3\hat{\bs{s}}_{\alpha i}\hat{\bs{S}}_{\alpha i+\hat{\bs{y}}} 
+ J'\hat{\bs{s}}_{3 i}\hat{\bs{S}}_{1 i+\hat{\bs{x}}}\right)\\
&+& \sum_{j\in{\cal B}}\left(\hat H^{\cal B}_j 
+ J\sum_{\alpha=1}^3 \hat{\bs{S}}_{\alpha j} \hat{\bs{s}}_{\alpha j+\hat{\bs{y}}} 
+ J'\hat{\bs{S}}_{3 j}\hat{\bs{s}}_{1 j+\hat{\bs{x}}}\right)\hspace{20pt}
\end{eqnarray}
When evaluating the spectrum of \eqref{eq:h2d} below, we will set
$J=1$ (and thereby equal to the intra-rung couplings).  The
inter-ladder coupling $J'$ is determined numerically by comparing the
energy difference of a $t$-$J$ model with a charge stripe with and
without frustrated boundary conditions to the corresponding difference
in a Heisenberg model, in which the charge stripe has been replaced by
an antiferromagnetic coupling $J'$ (see Appendix \ref{sec:jprime}); we
find $J'=0.07$.  For the moment, however, we keep the inter-rung and
inter-ladder couplings $J$ and $J'$ as free parameter, as this makes
it easier to trace the individual terms in the expansion below.

The next step is to expand \eqref{eq:h2d} in terms of our bosonic
creation and annihilation operators, using \eqref{eq:sa}, \eqref{eq:sb},
and
\begin{equation}\label{eq:spmzz}
  \bs{s}_\alpha\bs{S}_\beta =\frac{1}{2}
  \Big(\hat{s}_\alpha^+\hat{S}_\beta^- +\hat{s}_\alpha^-\hat{S}_\beta^+\Big)
  + \hat{s}_\alpha^z\hat{S}_\beta^z.
\end{equation}
We keep only terms up to second order in the operators.  Since
$\hat s_\alpha^z$ (and $\hat S_\alpha^z$) contains a constant term
with coefficient $s_{\alpha,11}^z$ as well as the linear terms
\begin{displaymath}
  s^z_{\alpha,13}\bigl(a_0^{\dagger}+a_0^{\phantom{\dagger}}\bigr) 
  + s^z_{\alpha,16}\bigl(c_0^{\dagger}+c_0\bigr),
\end{displaymath}
the expanded Hamiltonian will contain the linear term 
\begin{equation}\label{eq:linterm}
  \biggr(
  2J\sum_{\alpha=1}^3 S^z_{\alpha,11}s^z_{\alpha,16}
  +J'\big(S^z_{1,11}s^z_{3,16}+S^z_{3,11}s^z_{1,16}\big)\!
  \biggl)\bigl(c_{0}^{\dagger}+c_{0}\bigr)
\end{equation}
in addition to the linear term 
\begin{eqnarray}\label{eq:lintermha}
  \frac{3}{2}uv \bigl(c_{0}^{\dagger}+c_{0}\bigr)
\end{eqnarray}
already contained in \eqref{eq:ha} for each rung $i$ on sublattice
${\cal A}$.  The terms proportional to
$$\big(a_{0}^{\dagger}+a_{0}^{\phantom{\dagger}}\big)$$ cancel since
$s^z_{1,13}=-s^z_{3,13}$ and
$s^z_{2,13}=0$. 
This cancellation can also be inferred from symmetry considerations,
as elaborated in the following section.

We eliminate the linear terms \eqref{eq:linterm} and
\eqref{eq:lintermha} by adjusting the parameter $\phi$, \ie by solving
\begin{equation}\label{eq:lintermseq}
\frac{3}{2}uv+2J\sum_{\alpha=1}^3 S^z_{\alpha,11}s^z_{\alpha,16}+
2J'S^z_{1,11}s^z_{3,16} = 0
\end{equation}
with $J=1$, $J'=0.07$, $u=\cos\phi$, $v=\sin\phi$, and the matrix
elements $s^z_{\alpha,\mu\nu}$ 
as given in Appendix \ref{sec:spmz} in terms of $u$ and $v$.  This yields
\begin{eqnarray}\label{eq:phi}
\sqrt{2}\frac{v}{u} \approx 0.5019\quad\rm{or}\quad 
\phi = 0.3410.
\end{eqnarray}
On a formal level, the reason for rotating our basis states via
\eqref{eq:fidstate}, \eqref{eq:rota} and \eqref{eq:rotb} to begin with
was that this created the linear term \eqref{eq:lintermha} in $H^{\cal
  A}$.  Without this term, there would have been no way to eliminate
\eqref{eq:linterm}, and the basis set would have been highly
impractical for further analysis.  On a physical level, the
spontaneous breakdown of the SU(2) spin rotation symmetry leads us to
expect that the fiducial state of the rungs is much closer to the
classically ordered N\'eel state $\sket{\!\up\dw\up}$ than
$\sket{b_{-1/2}}$.  Not surprisingly, the spectrum evaluated below is
gapless at some point in the Brillouin zone, as required by
Goldstone's theorem for a state with a spontaneously broken continuous
symmetry, if and only if $\phi$ assumes the value \eqref{eq:phi}.

\begin{figure}[t]
  \begin{minipage}[c]{0.3\textwidth}
    \includegraphics[width=\linewidth]{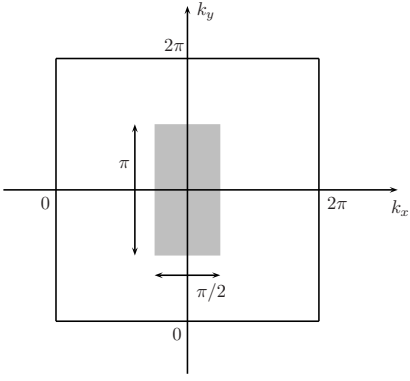}
  \end{minipage}
  \caption{The reduced Brillouin zone corresponding to the real space
    unit cell indicated in Fig.~\ref{fig:2dm} contains only 1/8th of
    the full Brillouin zone and is indicated by the shaded area.}
  \label{fig:bz}
\end{figure}

\section{Expanding the Hamiltonian}
\label{sec:expanding}

To evaluate the spectrum of \eqref{eq:h2d}, we first define
momentum space operators on sublattice $\cal A$ according to
\begin{eqnarray}\label{eq:fta}\nonumber
a_{0,\bk}&=&\sqrt{\frac{2}{N}}\sum_{i\in{\cal A}} e^{i\bk\bR_i}a_{0,i},\\
a_{0,i}&=&\sqrt{\frac{2}{N}}\sum_{\bk} e^{-i\bk\bR_i}a_{0,\bk},
\end{eqnarray}
where where $N$ denotes the number of rungs and the sums over $\bs{k}$
are taken over the reduced Brillouin zone indicated in gray in
Fig.~\ref{fig:bz}.  Similarly, for the creation and annihilation
operators on sublattice $\cal B$ we introduce
\begin{eqnarray}\label{eq:ftb}\nonumber
A_{0,\bk}&=&\sqrt{\frac{2}{N}}\sum_{j\in{\cal B}} e^{-i\bk\bR_j}A_{0,j},\\
A_{0,j}&=&\sqrt{\frac{2}{N}}\sum_{\bk} e^{i\bk\bR_j}A_{0,\bk},
\end{eqnarray}
which differ from \eqref{eq:fta} only in that the sign of the phases
is reversed.  Since we are only interested in the one-particle spin
wave spectrum, we neglect the effect of the Hilbert space restrictions
for the real space creation and annihilation operators (\ie that we
could create only one ``particle'' per rung) on the momentum space
operators.

As mentioned above, we only keep terms up to second order in the
creation and annihilation operators in the Hamiltonian.  When
substituting the explicit expressions \eqref{eq:sa} and \eqref{eq:sb}
into \eqref{eq:h2d}, we see immediately that many terms yield only
higher orders, while others cancel.  To begin with, since the spin
flip operators only contain terms of first and second order in the
creation and annihilation operators and are always multiplied with
another spin-flip operator, we only need to keep terms of first order
in the expressions for $\hat{s}_{\alpha}^{+}$ and
$\hat{S}_{\alpha}^{-}$.  The expansions of the
$\hat{s}_{\alpha}^z\hat{S}_{\beta}^z$ terms are slightly more
complicated, as $\hat{s}_{\alpha}^z$ and $\hat{S}_{\beta}^z$ contain
constant terms in addition to terms of first and second order in the
creation and annihilation operators.  
We have adjusted the parameter $\phi$ such that the linear terms in
the expansion cancel.  Most of the quadratic terms result from
multiplying the constant term $S^z_{\alpha,11}$ in the expansion of
$\hat{S}_{\alpha}^z$ with the quadratic terms in $\hat{s}_{\alpha}^z$
and multiplying $s^z_{\alpha,11}$ with quadratic terms in
$\hat{S}_{\alpha}^z$.  This yields seven diagonal terms (like
$b_{1}^{\dagger}b_{1}^{\phantom{\dagger}}$) and four off-diagonal
terms (like $b_1^{\dagger}c_1^{\phantom{\dagger}}
+c_{1}^{\dagger}b_{1}^{\phantom{\dagger}}$) for each sublattice.
Three of the off-diagonal terms, those linear in the antisymmetric
operators $a_{1}^{\dagger}$ or $a_{1}^{\phantom{\dagger}}$, vanish.
This can be seen either from the explicit coefficients written out in
Appendix \ref{sec:spmz} (\eg $s^z_{1,24}=-s^z_{3,24}$ and $s^z_{2,24}=0$
while $S^z_{1,11}=S^z_{3,11}$) or from a symmetry consideration.  As
the Hamiltonian is invariant under reflection symmetry interchanging
the outer chains of each three-leg ladder (\ie sites 1 and 3 on each
rung), there can only be terms containing an even number of the
antisymmetric operators $a_0^{\dagger}$, $a_0^{\phantom{\dagger}}$,
$a_1^{\dagger}$, $a_1^{\phantom{\dagger}}$, $A_{0}^{\dagger}$,
$A_{0}^{\phantom{\dagger}}$, $A_{-1}^{\dagger}$, or
$A_{-1}^{\phantom{\dagger}}$ in the expansion.

In addition to this reflection symmetry, we have the SU(2) spin
rotation symmetry of the Hamiltonian \eqref{eq:h2d}.  The spin
symmetry implies that the $z$-component of the total spin,
\begin{equation}
  \label{eq:stot}
  \hat{S}_{\text{tot}}^z = \sum_{i\in {\cal A}} \hat{s}_{\alpha i}^z
  + \sum_{j\in{\cal B}} \hat{S}_{\alpha j}^z,
\end{equation}
must be conserved.  This means that to second order in the creation
and annihilation operators, only operators which change $\hat{s}^z$ or
$\hat{S}^z$ by the same integer can appear in each term.  For
example, we can have a term $b_1^{\dagger}c_1^{\phantom{\dagger}}$ or
$b_1^{\dagger}c_{-1}^{\dagger}$, but not
$b_1^{\dagger}c_0^{\phantom{\dagger}}$.  Both symmetries together
imply that to second order, the Hamiltonian \eqref{eq:h2d} 
decomposes into terms which contain only operators belonging
to one particular group,
\begin{eqnarray}\label{eq:h2d-decompose}
\hat H = \tilde E_0+\hat H_{a0} + \hat H_{c0} + \hat H_{a1} 
+ \hat H_{c2} + \hat H_{b1,c1,c-1} ,
\end{eqnarray}
where $\tilde E_0$ is a contribution to the ground state energy, $\hat
H_{a0}$ contains only the operators $a_0^{\dagger}$,
$a_0^{\phantom{\dagger}}$, $A_{0}^{\dagger}$, and
$A_{0}^{\phantom{\dagger}}$,
and so on.  The low energy physics we are interested in is contained
in $\hat H_{b1,c1,c-1}$, which we will analyze in detail below.  As
for the other terms, explicit expressions and expansions in terms of
creation and annihilation operators are given in Appendix \ref{sec:other}.
$\hat H_{a0}$ and $\hat H_{c0}$ describe almost dispersionless modes
with energies of around $2.1$ and $2.7$ (in units of $J_{\text{exp}}$
which we eventually set to $J_{\text{exp}}=140\,\text{meV}$).  $\hat
H_{a1}$ describes a weakly dispersing mode of energy of about $2.0$,
with a bandwidth of about $0.2$.  $\hat H_{c_2}$ describes a completely
dispersionless mode with energy $3.12$.  Cuts of the dispersions of
these modes are shown in Fig.~\ref{fig:other} in Appendix \ref{sec:other}.
Since these modes occur at energies at which we consider our spin wave
theory no longer reliable, we will not discuss them further.

To evaluate the spectrum of $\hat H_{b1,c1,c-1}$, we write
\begin{eqnarray}\label{eq:hs}
\hat H_{b1,c1,c-1} = \sum_{\bk}\Big( 
\hat\Psi_{\bk}^{\dagger}H^{\phantom{\dagger}}_{\bk}\hat\Psi^{\phantom{\dagger}}_{\bk}
- \frac{1}{2}\text{tr}(H_{\bk})\Big),
\\[-12pt]\nonumber
\end{eqnarray}
where
\begin{eqnarray}\label{eq:psi_opdag}\nonumber
  \hat\Psi^{\dagger}_{\bk} &=& 
  \bigl( B^{\dagger}_{-1,\bk},  b_{1,\bk}^{\phantom{\dagger}}, 
  C^{\dagger}_{-1,\bk}, c_{1,\bk}^{\phantom{\dagger}}, 
  C_{1,-\bk}^{\phantom{\dagger}}, c^{\dagger}_{-1,-\bk}\bigr),\\\label{eq:psi_op}
  \hat\Psi^{\phantom{\dagger}}_{\bk} &=& 
  \bigl( B_{-1,\bk}^{\phantom{\dagger}},  b^{\dagger}_{1,\bk}, 
  C_{-1,\bk}^{\phantom{\dagger}}, c^{\dagger}_{1,\bk},  
  C^{\dagger}_{1,-\bk}, c_{-1,-\bk}^{\phantom{\dagger}}\bigr)^{\rm T}.\hspace{10pt}
\end{eqnarray}
The $6\times 6$ matrix $H_{\bk}$ consists of the $\bs{k}$-independent 
diagonal terms 
\begin{eqnarray}\label{eq:h11}\nonumber
  H_{\bk,11}=H_{\bk, 22} \!&=&\!
  -\frac{3}{2}v^2 
  +2J\sum_{\alpha=1}^3 s^z_{\alpha,11}(S^z_{\alpha,22}-S^z_{\alpha,11})
  \\[4pt]\nonumber
  &+&\! 2J'\,s^z_{3,11}(S^z_{1,22}-S^z_{1,11}),
  \\\nonumber
  H_{\bk,33}=H_{\bk, 44} \!&=&\!
  \frac{3}{2}u^2
  + 2J\sum_{\alpha=1}^3s^z_{\alpha,11}(S^z_{\alpha,77}-S^z_{\alpha,11})
  \\\nonumber
  &+&\! 2J'\,s^z_{3,11}(S^z_{1,77}-S^z_{1,11}),
  \\\nonumber
  H_{\bk,55}=H_{\bk, 66} \!&=&\! 
  \frac{3}{2}u^2 
  + 2J\sum_{\alpha=1}^3 s^z_{\alpha,11}(S^z_{\alpha,55}-S^z_{\alpha,11})
  \\ 
  &+&\! 2J'\, s^z_{3,11}(S^z_{1,55}-S^z_{1,11}).
\end{eqnarray}
The off-diagonal terms are of the general form
\begin{equation}\nonumber
H_{\bk,ij} = H_{\bk,ji} = H_{ij}^0 + H_{ij}^x\,\cos(4k_x) + H_{ij}^y\,\cos(k_y).
\end{equation}
The $\bs{k}$-independent coefficients
\begin{equation}\label{eq:h15} 
H^0_{13}=H^0_{24}=2J\sum\limits_{\alpha=1}^3 s^z_{\alpha,11}S^z_{\alpha,27}
+2J's^z_{3,11}S^z_{1,27}\hspace{10pt} 
\end{equation}
result from the $\hat{s}_{\alpha}^z\hat{S}_{\beta}^z$ terms.
Expansion of the $\hat{s}_{\alpha}^+\hat{S}_{\beta}^-$ 
terms
yields the coefficients
\begin{eqnarray}\label{eq:hx} 
\begin{array}{l@{\,}c@{\,}r@{\qquad}l@{\,}c@{\,}r}
H_{12}^x &=& J' s^+_{3,21}S^-_{1,21}, &
H_{14}^x &=& J' s^+_{3,71}S^-_{1,21}, \\[1mm]
H_{16}^x &=& J' s^+_{3,15}S^-_{1,21}, &
H_{23}^x &=& J' s^-_{3,21}S^+_{1,71}, \\[1mm]
H_{25}^x &=& J' s^-_{3,21}S^+_{1,51}, &
H_{34}^x &=& J' s^+_{3,71}S^-_{1,71}, \\[1mm]
H_{36}^x &=& J' s^+_{3,15}S^-_{1,71}, &
H_{45}^x &=& J' s^-_{3,71}S^+_{1,51}, \\[1mm]
H_{56}^x &=& J' s^+_{3,15}S^-_{1,15}, 
\end{array}
\end{eqnarray}
and
\begin{eqnarray}\label{eq:hy} 
\begin{array}{l@{\,}c@{\,}r@{\quad}l@{\,}c@{\,}r}
H^y_{12} &=& J\sum\limits_{\alpha=1}^3 s^+_{\alpha,21}S^-_{\alpha,21},&
H^y_{14} &=& J\sum\limits_{\alpha=1}^3 s^+_{\alpha,71}S^-_{\alpha,21},\\[1mm] 
H^y_{16} &=& J\sum\limits_{\alpha=1}^3 s^+_{\alpha,15}S^-_{\alpha,21},&        
H^y_{23} &=& J\sum\limits_{\alpha=1}^3 s^-_{\alpha,12}S^+_{\alpha,17},\\[1mm]    
H^y_{25} &=& J\sum\limits_{\alpha=1}^3 s^+_{\alpha,21}S^-_{\alpha,15},&
H^y_{34} &=& J\sum\limits_{\alpha=1}^3 s^+_{\alpha,71}S^-_{\alpha,71},\\[1mm]
H^y_{36} &=& J\sum\limits_{\alpha=1}^3 s^+_{\alpha,15}S^-_{\alpha,71},&
H^y_{45} &=& J\sum\limits_{\alpha=1}^3 s^-_{\alpha,17}S^+_{\alpha,51},\\[1mm]
H^y_{56} &=& J\sum\limits_{\alpha=1}^3 s^+_{\alpha,15}S^-_{\alpha,15}.           
\end{array}
\end{eqnarray}
All other off-diagonal elements of $H_{\bs{k},ij}$ vanish. 


\section{Solution by Bogoliubov transformation}
\label{sec:solution}

The Hamiltonian \eqref{eq:hs} can be diagonalized 
with a $2n$ dimensional Bogoliubov transformation
\cite{sachdev92prb12377}.  We begin with a brief review of the
formalism.

At each point in $\bs{k}$-space, we wish to write the 
Hamiltonian in terms of a diagonal matrix $\Omega$,
\begin{equation}
  \label{eq:bogham}
  \hat H =\hat\Psi^{\dagger} H \hat\Psi 
= \hat\Gamma^{\dagger}\Omega\hat\Gamma,
\end{equation}
with 
\begin{equation}
  \label{eq:defofm}
  \hat\Psi = M\hat\Gamma ,\qquad \Omega =M^{\dagger} H M .
\end{equation}
The components of $\hat\Gamma$ satisfy the same commutation relations as
the components of $\hat\Psi$:
\begin{equation}
  \label{eq:comrel}
  \big[\hat\Psi_i, \hat\Psi_j^{\dagger}\big] 
  = \big[\hat\Gamma_i, \hat\Gamma_j^{\dagger}\big] = T_{ij}
\end{equation}
with 
\begin{equation}
T = \mathrm{diag}\left(1,-1,1,-1,-1,1\right).
\end{equation}
This implies
\begin{multline}\nonumber
\hspace{10pt} T_{ij} = \big[\hat\Psi_i,\hat\Psi_j^{\dagger}\big]
= \sum_{l,m} \big[ M_{il}\hat\Gamma_l, \hat\Gamma^{\dagger}_m (M^{\dagger})_{mj} \big]
\\\nonumber 
= \sum_{l,m} M_{il}\big[ \hat\Gamma_l, \hat\Gamma_m^{\dagger}\big] M^{\dagger}_{mj} 
= \sum_{l,m} M_{il}T_{lm}M^{\dagger}_{mj},\hspace{10pt}
\end{multline}
or
\begin{equation}
  \label{eq:tmtm}
  T=MTM^{\dagger}.
\end{equation}
Multiplying \eqref{eq:tmtm} from the right by $HM$ yields with
\eqref{eq:defofm}
\begin{equation}
  \label{eq:m_as_ev}
  T H M = M T\Omega ,
\end{equation}
or in components
\begin{equation}
  \label{eq:m_as_evc}
  \sum_{l} (T H)_{il} M_{lj}=  M_{ij} (T\Omega)_{jj},
\end{equation}
\ie the $j$-th column of $M$ is given by an eigenvector of $T H$ with
eigenvalue $T_{jj}\Omega_{jj}$. 
This specifies $M$ up to the normalization of the
eigenvectors.  To obtain the normalization, it is propitious to
rewrite \eqref{eq:tmtm} as
\begin{equation}
  \label{eq:mtmt}
  T=M^{\dagger}TM.
\end{equation}
(To obtain \eqref{eq:mtmt}, multiply \eqref{eq:tmtm} by $TM^{-1}$ from
the left and by $TM$ from the right and use $T^2=1$.)  Each column $j$
of $M_{ij}$ must hence be normalized such that
\begin{equation}
  \label{eq:norm}
  T_{jj} =\sum_i T_{ii} |M_{ij}|^2.
\end{equation}
\begin{figure}[t]
  \begin{center}
    \includegraphics[height=65mm]{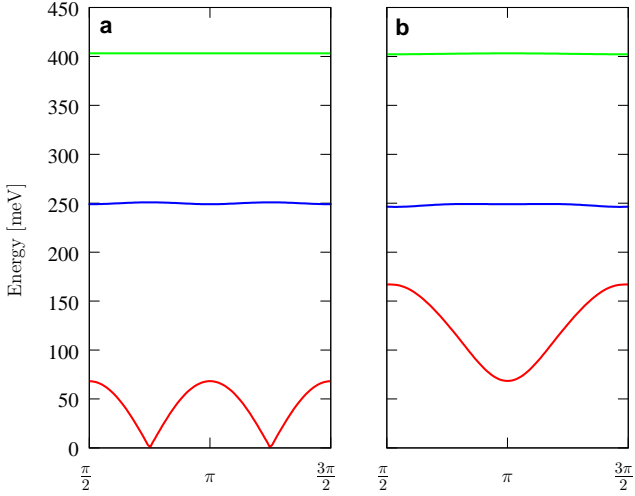}
  \end{center}
  \caption{(Color online)\ Modes described by $\hat H_{b1,c1,c-1}$
    plotted as cuts (a) along $(k_x,\pi)$ and (b) along $(\pi,k_y)$
    using $J_{\text{exp}}=140\,\text{meV}$.}
  \label{fig:h_bcc}
\end{figure}

\begin{figure}[t]
    \includegraphics[width=\linewidth]{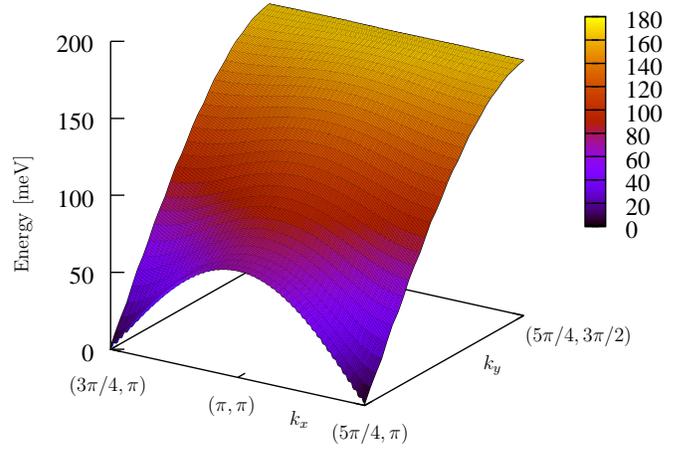}
    \caption{(Color online) The dispersion $\omega(k_x,k_y)$ of the
      lowest eigenmode of $\hat H_{b1,c1,c-1}$ in half of the reduced
      Brillouin assuming $J_{\text{exp.}}=140\,\rm{meV}$}
    \label{fig:plot_3d}
\end{figure}

Diagonalization of $\eqref{eq:hs}$ using this formalism at
each point in $\bs{k}$ space with 
\begin{eqnarray}\label{eq:gamma_ops}\nonumber
  \hat\Gamma_{\bk}^{\dagger}&=&\bigl(
  \gamma_{1,\bk}^{\dagger},\gamma_{2,\bk}^{\phantom{\dagger}},
  \gamma_{3,\bk}^{\dagger},\gamma_{4,\bk}^{\phantom{\dagger}},
  \gamma_{5,\bk}^{\phantom{\dagger}},\gamma_{6,\bk}^{\dagger}
  \bigr),\\
  \hat\Gamma_{\bk}^{\phantom{\dagger}}&=&\bigl(
  \gamma_{1,\bk}^{\phantom{\dagger}},\gamma_{2,\bk}^{\dagger},
  \gamma_{3,\bk}^{\phantom{\dagger}},\gamma_{4,\bk}^{\dagger},
  \gamma_{5,\bk}^{\dagger},\gamma_{6,\bk}^{\phantom{\dagger}}
  \bigr)^{\rm T},\hspace{10pt}
\end{eqnarray}
yields 
\begin{equation}\label{eq:hgamma}
  \hat H_{b1,c1,c-1} = \sum_{\bk,i}\left[\,\omega_{\bk,i}^{\phantom{\dagger}}\,
  \gamma^{\dagger}_{i,\bk}\gamma_{i,\bk}^{\phantom{\dagger}}
  + \frac{1}{2}\big(
  \omega_{\bk,i}^{\phantom{\dagger}}-H_{\bk, ii}^{\phantom{\dagger}}
  \big)\right].
\end{equation}
This Hamiltonian describes three two-fold degenerate modes
$\omega_{\bk,i}$, which we have plotted assuming $J_{\text{exp.}}=140\,
\text{meV}$ as cuts along $(k_x,\pi)$ and $(\pi,k_y)$ in
Fig.~\ref{fig:h_bcc}.  The two-fold degeneracy of each mode
corresponds to spin waves with $S_z=\pm 1$.  Since we expect our spin
wave theory to be reliable only for energies up to $J_{\text{exp.}}$,
we will disregard the higher modes along with those analyzed in
Appendix \ref{sec:other}.  The lowest mode
$\omega_{\bk,1}=\omega_{\bk,2}:=\omega (\bk)$ is shown as a 3D plot
for half of the reduced Brillouin zone in Fig.~\ref{fig:plot_3d}.

The Hamiltonian \eqref{eq:hgamma} further contains a contribution
\begin{equation}
  \label{eq:b1c1c-1_gsenergy}
  E_{b1,c1,c-1} = 
  \sum_{\bk,i}\frac{1}{2}\big(
  \omega_{\bk,i}^{\phantom{\dagger}}-H_{\bk, ii}^{\phantom{\dagger}}
  \big) = 
  -0.22116\,N 
\end{equation}
to the ground state energy.  (Here $N$ denotes the number of rungs.
The sum extends over $\frac{N}{2}$ values for $\bk$.)
In Appendix \ref{sec:groundstate}, we obtain the ground state energy
by adding this contribution to the contributions of the other terms in
\eqref{eq:h2d-decompose} given in Appendix \ref{sec:other}, and obtain
$E_0=-1.73378\,N$.  We find that this number is in excellent agreement
with what we would expect from diagonalizing the model for a finite
cluster with unfrustrated boundary conditions.  This confirms the
validity of our analysis.


\section{Discussion of the results}
\label{sec:discussion}

\subsection{Agreement with the experimental data}

The significance of our results emerges in the context of a comparison
of our spectrum with the experimental data obtained by Tranquada \ea
~\cite{tranquada-04n534} through inelastic neutron scattering on the
stripe ordered compound La$_{1.875}$Ba$_{0.125}$CuO$_{4}$.  The data
points with their corresponding error bars are shown as black or blue
crosses in Figs.~\ref{fig:tddtr}a and \ref{fig:tddtr}b, respectively.

%

In Fig.~\ref{fig:tddtr}b, which is directly reproduced from Tranquada
\ea\cite{tranquada-04n534}, the neutron data are superposed with the
spectrum of the triplet excitation of an isotropic two-leg Heisenberg
ladder, which models bond centered stripes at accordingly high
energies.  In Fig.~\ref{fig:tddtr}a, we have superposed cuts of the
lowest mode $\omega (\bs{k})$ along $(k_x,\pi)$ and $(\pi,k_y)$ with
the experimental data.  (The superposition of cuts of our spectrum in
the $x$ and $y$ direction reflects the assumption that a superposition
of domains with stripes along the two principal lattice directions has
been observed in the experiment.)  We believe it is fair to say that
up to energies of about $180\,\text{meV}$, the agreement is excellent.
(Above these energies, or more precisely above energies of order $J$,
a perturbative spin wave analysis becomes unreliable.)  Likewise, the
constant energy slices of the neutron scattering intensities
$\chi^{+-}(\bk,\omega)$ obtained with the matrix elements calculated
in Appendix \ref{sec:matrixelements} shown in Fig.~\ref{fig:neutron}
agree very well with the experimentally measured constant-energy
slices of the magnetic scattering in La$_{1.875}$Ba$_{0.125}$CuO$_{4}$
shown in Fig.~2 of Tranquada \ea\cite{tranquada-04n534}.

\begin{figure}[t]
  \includegraphics[width=\linewidth]{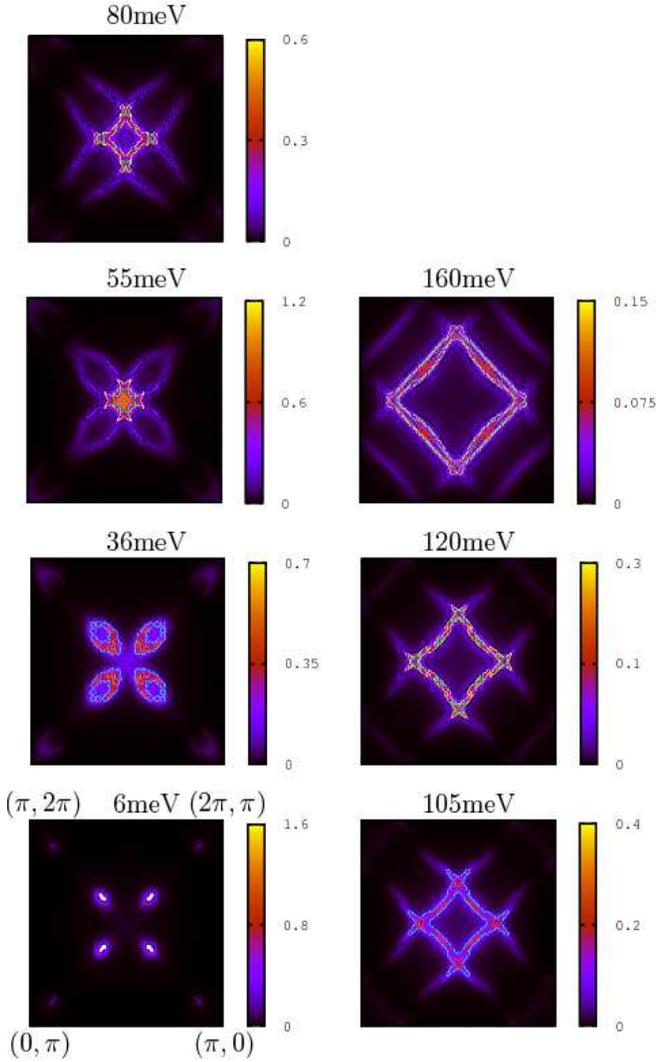}
  \caption{(Color online) Constant energy slices of the neutron
    scattering intensity $\chi^{+-}(\bk,\omega)$ (see Appendix
    \ref{sec:matrixelements} below) for $J_\text{exp}=140\,\text{meV}$
    and $J'=0.07\,J_\text{exp}$ in the magnetic Brillouin zone.  In
    the indicated energy range, only the lowest mode shown in Figs.\
    \ref{fig:plot_3d} and \ref{fig:tddtr}a contributes.  We have
    replaced the $\delta$-functions in frequency by Lorentzians with
    half-width $\Delta=0.05\,J_\text{exp}$ and averaged over both
    stripe orientations (\ie horizontal and vertical).}
  \label{fig:neutron}
\end{figure}

\begin{figure}[t]
  \begin{center}
\includegraphics[width=0.8\linewidth]{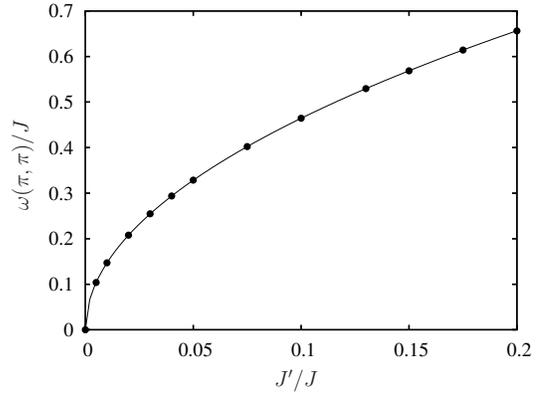}
  \end{center}
  \vspace{-10pt}
  \caption{The saddle point energy $\omega(\pi,\pi)$ of the lowest
    magnetic mode $\omega (\bk)$ calculated for various values of
    $J'/J$ (points).  Fitting yields $\omega(\pi,\pi)\approx
    1.47\sqrt{J'J}$ (solid line).}
  \label{fig:ej}
\end{figure}

\subsection{Dependence on the inter-ladder coupling $\bs{J'}$}

The good agreement of our results with the experimental data up to
energies even larger than $J=140\,\text{meV}$ (where we would expect
that the perturbative spin wave analysis becomes unreliable) is
somewhat surprising.  While any explanation in terms of bond-centered
stripes through coupled two-leg
ladders\cite{tranquada-04n534,vojta-04prl127002, uhrig-04prl267003}
gives immediately a roughly adequate estimate for the saddle-point
energy in terms of the triplet energy gap of the individual two-leg
ladders, it is far from obvious that a model of coupled three-leg
ladders, which are individually gapless, should give a saddle-point
energy consistent with the data.  In our model, the saddle-point
energy depends significantly on the coupling $J'$ between the ladders.
Fortunately, however, it is possible to determine $J'$ rather
accurately through numerical comparison of a $t$--$J$ model with a
site centered spin and a site centered charge stripe to a model with
three-leg Heisenberg ladders coupled by $J'$, as described in Appendix
\ref{sec:jprime}.  This analysis does not only provide us with the
value $J'\approx 0.07J$, but also shows that this value is rather
robust in the sense that it does not significantly depend on the
details of how we localize the stripe.  To obtain a better
understanding of the dependence of our final results on this coupling,
we have obtained the spectrum for a number of different values of $J'$
by solving \eqref{eq:lintermseq} numerically for each value, and
proceeding with the Bogoliubov transformation with the resulting
values for $u(J')$ and $v(J')$.  The results for the saddle-point
energies $\omega(\pi,\pi)$ are shown in Fig.~\ref{fig:ej} (black
dots).  Fitting the data yields
\begin{equation}
  \label{eq:jprimefit}
  \omega(\pi,\pi)\approx 1.47 \sqrt{J'J}  
\end{equation}
to an excellent approximation up to values where $J'$ becomes
comparable to $J$ (solid line).

\begin{figure}[t]
  \begin{center}
    \includegraphics[width=0.96\linewidth]{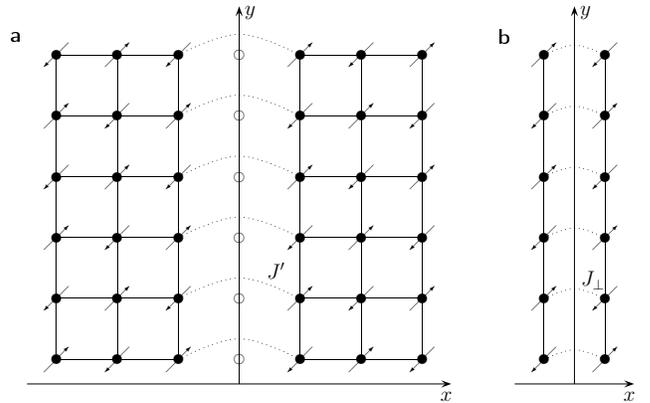}
  \end{center}
  \vspace{-6pt}
  \caption{Auxiliary models of (a) two weakly coupled three-leg
    ladders and (b) two weakly coupled spin \half chains used in the
    discussion to understand the square root dependence of
    $\omega(\pi,\pi)$ on $J'$ depicted in Fig.~\ref{fig:ej}.}
  \label{fig:ladders}
\end{figure}
The square root dependence of $\omega(\pi,\pi)$ on $J'$ can be
understood by considering a model of two three-leg ladders, which are
weakly coupled by $J'$, as shown in Fig.~\ref{fig:ladders}a. The
low energy excitations of the individual three-leg ladders are spin
\half spinons, which are gapless.  The coupling $J'$ induces a linear
confinement potential  
\begin{equation}
  \label{eq:linconf}
  V(y)=F|y|
\end{equation}
between pairs of spinons, since the links coupling the chains
effectively become decorrelated in the region between them.  The
situation here is similar to a system of two coupled spin \half chains
shown in Fig.~\ref{fig:ladders}b, where a weak coupling $J_{\perp}$
between the chains is known to induce a linear confinement potential
between pairs of spinons \cite{shelton-96prb8521}.  In the model of
coupled chains, the force between the spinons is proportional
to\cite{greiter02prb134443,greiter02prb054505}
\begin{equation}
  \label{eq:fchains}
  F\approx\langle\bs{S}\bs{S}\rangle_{\perp} J_{\perp}\propto J_{\perp}^2/J.  
\end{equation}
For the model of two coupled three-leg ladders we consider here in the
context of understanding the dependence \eqref{eq:jprimefit} of our
spin wave analysis, however, we assume that the spin correlation
$\langle\bs{S}\bs{S}\rangle$ between the sites coupled by $J'$ is to
lowest order independent of $J'/J$.  Therewith we account for the
static correlations present due to the long range order we assume.
For the confinement force in our auxiliary model of two coupled
three-leg ladders we hence assume
\begin{equation}
  \label{eq:fladders}
  F\propto   J'.  
\end{equation}
The spinon confinement will then induce a gap $\Delta$, which
corresponds to the the ground state or zero-point energy of the linear
potential oscillator for the relative motion of the spinons.
The dispersion of the spinons is linear for both the individual spin
chains and the individual three-leg ladders\cite{rice-97prb14655},
\begin{equation}
  \label{eq:lindispers}
  \varepsilon(k_y)\approx v|k_y|,
\end{equation}
with $v\propto J$ as $J$ is the only energy scale is these models.
(In \eqref{eq:lindispers}, we have shifted both spinon branches to the
origin.)  The ground state energy $E_0$ of a constant force $F$
oscillator of linearly dispersing particles with velocity $v$,
however, is proportional to $\sqrt{Fv}$~~\cite{greiter10ap1349}.  
%
%
This implies $\Delta\propto J_{\perp}$ for the two weakly coupled
chains and $\Delta\propto \sqrt{J'J}$ for the two weakly coupled
three-leg ladders with the additional assumption of static
correlations sue to long range order.

To see why this gap $\Delta$ corresponds to the saddle point energy
$\omega(\pi,\pi)$ in the spin wave analysis above, consider the
transformation properties of our auxiliary model of the two weakly
coupled three-leg ladders shown in Fig.~\ref{fig:ladders}a under the
parity reflection $x\rightarrow -x$.  The gapped spinon-spinon bound
state is odd under this symmetry, which in the language of momenta
$k_x$ of the site centered stripe model corresponds to a shift of
$\frac{\pi}{4}$.  Since the ground state of the stripe model has order
with $k_x=\pi\pm \frac{\pi}{4}$, the gapped excitation will correspond
to $k_x=\pi$.  Following this line of reasoning, we can 
understand the square root dependence \eqref{eq:jprimefit} of
$\omega(\pi,\pi)$ depicted in Fig.~\ref{fig:ej}.
 
\section{Conclusions}
\label{sec:conclusions}

In this work, we have provided a full and detailed account of a spin
wave analysis of a coupled three-leg ladder model for spin stripes in
copper oxide superconductors.  We have numerically evaluated the
inter-ladder coupling $J'$ induced by the charge stripes in between
both site-centered spin stripes modelled by three-leg ladders and
bond-centered spin stripes modelled by two-leg ladders.  As reported
previously~\cite{greiter-10prb144509}, for the latter we obtain a
ferromagnetic coupling $J'=-0.05J$, which is not sufficient to close
the energy gap of the individual two-leg ladders.  This does not imply
that the stripes cannot be bond-centered.  It does imply, however,
that a description in terms of spin-only models of coupled two-leg
ladders is not sensible.

For site-centered spin stripes modelled by three-leg ladders, we
obtain an antiferromagnetic coupling $J'=0.07J$.  We have calculated
the spectrum, the staggered magnetization, the dynamical structure
factor $\chi^{+-}(\bk,\omega)$, and the ground state energy for a
spin-only 
model of coupled three-leg ladders using a linear spin wave
analysis of bosonic operators representing the eight-dimensional
Hilbert spaces on each three-site rung.  The analysis makes no
assumptions except for the model itself, and contains no variational
parameter, as even the inter-ladder coupling $J'$ is evaluated through
exact diagonalizations of small $t$--$J$ clusters with and without
charge stripes in between the spin stripes.  We find excellent
agreement with the experimental data of Tranquada
\ea~\cite{tranquada-04n534}.  The experimental data hence point
towards site-centered, and not, as previously asserted,
bond-centered~\cite{tranquada-04n534,vojta-04prl127002,uhrig-04prl267003}
stripes. \\[-12pt]


\section*{ACKNOWLEDGMENTS}

We wish to thank Matthias Vojta, Tobias Ulbricht, and Peter W\"olfle
for discussions of this work.  This work was supported by the German
Research Foundation under grant FOR 960.



\appendix

\section{Matrix elements of the individual spin operators on rungs}
\label{sec:spmz}

In Sec.~\ref{sec:basis}, we have written out the spin operators
$\hat{s}_{\alpha}^{\tau}$ and $\hat{S}_{\alpha}^{\tau}$ with
$\tau=+,-,z$ on the individual sites $\alpha=1,2,3$ on rungs belonging
to sublattice $\cal A$ and $\cal B$, respectively,
\begin{eqnarray}
  \hat{s}_\alpha^{\tau}&=&
  \sum_{\mu,\nu}s_{\alpha,\mu\nu}^{\tau}\ket{\mu}\sbra{\nu}^{\cal A},\\
  \hat{S}_\alpha^{\tau}&=& 
  \sum_{\mu,\nu}S_{\alpha,\mu\nu}^{\tau}\ket{\mu}\sbra{\nu}^{\cal B},
\end{eqnarray}
in the basis sets $M^{\cal A}$ and $M^{\cal B}$ specified in \eqref{eq:ma} and
\eqref{eq:mb}.  The matrix elements
\begin{eqnarray}
  s_{\alpha,\mu\nu}^{\tau}&=&\sbra{\mu}\hat{s}_{\alpha}^{\tau}\ket{\nu}^{\cal A}
  \\
  S_{\alpha,\mu\nu}^{\tau}&=&\sbra{\mu}\hat{S}_{\alpha}^{\tau}\ket{\nu}^{\cal B},
  \label{eq:matrixab}
\end{eqnarray}
are explicitly given by:
\begin{widetext}
\begin{eqnarray}
s_1^+ &=&
\begin{pmatrix}
0 & 0 & 0 & 0 & \frac{-u+\sqrt{2}v}{\sqrt{6}} & 0 & 0 & 0 \\
\frac{-2\sqrt{2}u+v}{3\sqrt{2}} & 0 & -\frac{1}{\sqrt{3}} & 0 & 0 & \frac{u+\sqrt{2}v}{3\sqrt{2}} & 0 & 0 \\
0 & 0 & 0 & 0 & \frac{1}{\sqrt{2}} & 0 & 0 & 0 \\
-\frac{\sqrt{2}u+v}{\sqrt{6}} & 0 & 0 & 0 & 0 & \frac{-u+\sqrt{2}v}{\sqrt{6}} & 0 & 0 \\
0 & 0 & 0 & 0 & 0 & 0 & 0 & 0 \\
0 & 0 & 0 & 0 & \frac{\sqrt{2}u+v}{\sqrt{6}} & 0 & 0 & 0 \\
\frac{u+2\sqrt{2}v}{3\sqrt{2}} & 0 & -\frac{1}{\sqrt{6}} & 0 & 0 & \frac{2\sqrt{2}u-v}{3\sqrt{2}} & 0 & 0 \\
0 & -\frac{1}{\sqrt{6}} & 0 & \frac{1}{\sqrt{2}} & 0 & 0 & \frac{1}{\sqrt{3}} & 0 \\
\end{pmatrix}
= \left(s_1^-\right)^{\dagger} = S_1^- = \left(S_1^+\right)^{\dagger}
\\[15pt]
%
s_2^+ &=&
\begin{pmatrix}
0 & 0 & 0 & 0 & \frac{\sqrt{2}u+v}{\sqrt{3}} & 0 & 0 & 0 \\
\frac{u-\sqrt{2}v}{3} & 0 & 0 & 0 & 0 & -\frac{\sqrt{2}u+v}{3} & 0 & 0 \\
0 & 0 & 0 & 0 & 0 & 0 & 0 & 0 \\
0 & 0 & -1 & 0 & 0 & 0 & 0 & 0 \\
0 & 0 & 0 & 0 & 0 & 0 & 0 & 0 \\
0 & 0 & 0 & 0 & \frac{u-\sqrt{2}v}{\sqrt{3}} & 0 & 0 & 0 \\
\frac{-\sqrt{2}u+2v}{3} & 0 & 0 & 0 & 0 & \frac{2u+\sqrt{2}v}{3} & 0 & 0 \\
0 & \sqrt{\frac{2}{3}} & 0 & 0 & 0 & 0 & \frac{1}{\sqrt{3}} & 0 
\end{pmatrix}
= \left(s_2^-\right)^{\dagger} = S_2^- = \left(S_2^+\right)^{\dagger}
\\[15pt]
%
s_3^+ &=& 
\begin{pmatrix}
0 & 0 & 0 & 0 & \frac{-u+\sqrt{2}v}{\sqrt{6}} & 0 & 0 & 0 \\
\frac{-2\sqrt{2}u+v}{3\sqrt{2}} & 0 & \frac{1}{\sqrt{3}} & 0 & 0 & \frac{u+\sqrt{2}v}{3\sqrt{2}} & 0 & 0 \\
0 & 0 & 0 & 0 & -\frac{1}{\sqrt{2}} & 0 & 0 & 0 \\
\frac{\sqrt{2}u+v}{\sqrt{6}} & 0 & 0 & 0 & 0 & \frac{u-\sqrt{2}v}{\sqrt{6}} & 0 & 0 \\
0 & 0 & 0 & 0 & 0 & 0 & 0 & 0 \\
0 & 0 & 0 & 0 & \frac{\sqrt{2}u+v}{\sqrt{6}} & 0 & 0 & 0 \\
\frac{u+2\sqrt{2}v}{3\sqrt{2}} & 0 & \frac{1}{\sqrt{6}} & 0 & 0 & \frac{2\sqrt{2}u-v}{3\sqrt{2}} & 0 & 0 \\
0 & -\frac{1}{\sqrt{6}} & 0 & -\frac{1}{\sqrt{2}} & 0 & 0 & \frac{1}{\sqrt{3}} & 0 \\
\end{pmatrix}
= \left(s_3^-\right)^{\dagger} = S_3^- = \left(S_3^+\right)^{\dagger}
\\[15pt]
%
\label{eq:sz1} 
s^z_1 &=& 
\begin{pmatrix}
-\frac{\left(\sqrt{2}u+v\right)^2}{6} & 0 & \frac{-u+\sqrt{2}v}{2\sqrt{3}} & 0 & 0 & 
\frac{-\sqrt{2}u^2+uv+\sqrt{2}v^2}{6}& 0 & 0 \\
0 & \frac{1}{3} & 0 & \frac{1}{2\sqrt{3}} & 0 &0 & \frac{1}{3\sqrt{2}} & 0 \\
\frac{-u+\sqrt{2}v}{2\sqrt{3}} & 0 & 0 & 0 & 0 & \frac{\sqrt{2}u+v}{2\sqrt{3}} & 0 & 0\\
0 & \frac{1}{3\sqrt{2}} & 0 & 0 & 0 & 0 & -\frac{1}{\sqrt{6}} & 0 \\
0 & 0 & 0 & 0 & -\frac{1}{2} & 0 & 0 & 0 \\
\frac{-\sqrt{2}u^2+uv+\sqrt{2}v^2}{6} & 0 &  \frac{\sqrt{2}u+v}{2\sqrt{3}} &
 0 & 0 & -\frac{\left(u-\sqrt{2}v\right)^2}{6} & 0 & 0 \\
0 & \frac{1}{3\sqrt{2}} & 0 & -\frac{1}{\sqrt{6}} & 0 & 0 & \frac{1}{6} & 0 \\
0 & 0 & 0 & 0 & 0 & 0 & 0 &  \frac{1}{2} 
\end{pmatrix}
=-S^z_1
\\[15pt]
%
\label{eq:sz2} 
s^z_2 &=& 
\begin{pmatrix}
\frac{u^2+4\sqrt{2}uv-v^2}{6} & 0 & 0 & 0 & 0 & \frac{\sqrt{2}u^2-uv-\sqrt{2}v^2}{3}  & 0 & 0 \\
0 & -\frac{1}{6} & 0 & 0 & 0 & 0 & -\frac{\sqrt{2}}{3} & 0 \\
0 & 0 & -\frac{1}{2} & 0 & 0 & 0 & 0 & 0 & \\
0 & 0 & 0 & \frac{1}{2} & 0 & 0 & 0 & 0 \\
0 & 0 & 0 & 0 &  -\frac{1}{2} & 0 & 0 & 0 \\
\frac{\sqrt{2}u^2-uv-\sqrt{2}v^2}{3}& 0 & 0 & 0 & 0 &  \frac{-u^2-4\sqrt{2}uv+v^2}{6}& 0 & 0 \\
0 & -\frac{\sqrt{2}}{3} & 0 & 0 & 0 & 0 & \frac{1}{6} & 0 \\
0 & 0 & 0 & 0 & 0 & 0 & 0 & \frac{1}{2}  
\end{pmatrix}
=-S^z_2
\\[15pt]
%
\label{eq:sz3} 
s^z_3 &=& 
\begin{pmatrix}
-\frac{\left(\sqrt{2}u+v\right)^2}{6} & 0 & \frac{u-\sqrt{2}v}{2\sqrt{3}} & 0 & 0 & 
\frac{-\sqrt{2}u^2+uv+\sqrt{2}v^2}{6} & 0 & 0 \\
0 & \frac{1}{3} & 0 & -\frac{1}{2\sqrt{3}} & 0 &0 & \frac{1}{3\sqrt{2}} & 0 \\
\frac{u-\sqrt{2}v}{2\sqrt{3}} & 0 & 0 & 0 & 0 & -\frac{\sqrt{2}u+v}{2\sqrt{3}} & 0 & 0\\
0 & -\frac{1}{3\sqrt{2}} & 0 & 0 & 0 & 0 & \frac{1}{\sqrt{6}} & 0 \\
0 & 0 & 0 & 0 & -\frac{1}{2} & 0 & 0 & 0 \\
\frac{-\sqrt{2}u^2+uv+\sqrt{2}v^2}{6} & 0 & -\frac{\sqrt{2}u+v}{2\sqrt{3}} & 0 & 0 & 
-\frac{\left(u-\sqrt{2}v\right)^2}{6} & 0 & 0 \\
0 & \frac{1}{3\sqrt{2}} & 0 & \frac{1}{\sqrt{6}} & 0 & 0 & \frac{1}{6} & 0 \\
0 & 0 & 0 & 0 & 0 & 0 & 0 &  \frac{1}{2} 
\end{pmatrix}
=-S^z_3
\end{eqnarray}

\end{widetext}
\setlength{\columnwidth}{\linewidth}


\section{Estimation of the coupling $\bs{J'}$ between ladders}
\label{sec:jprime}

\begin{figure}[t]
  \begin{center}
    \includegraphics[width=0.90\columnwidth]{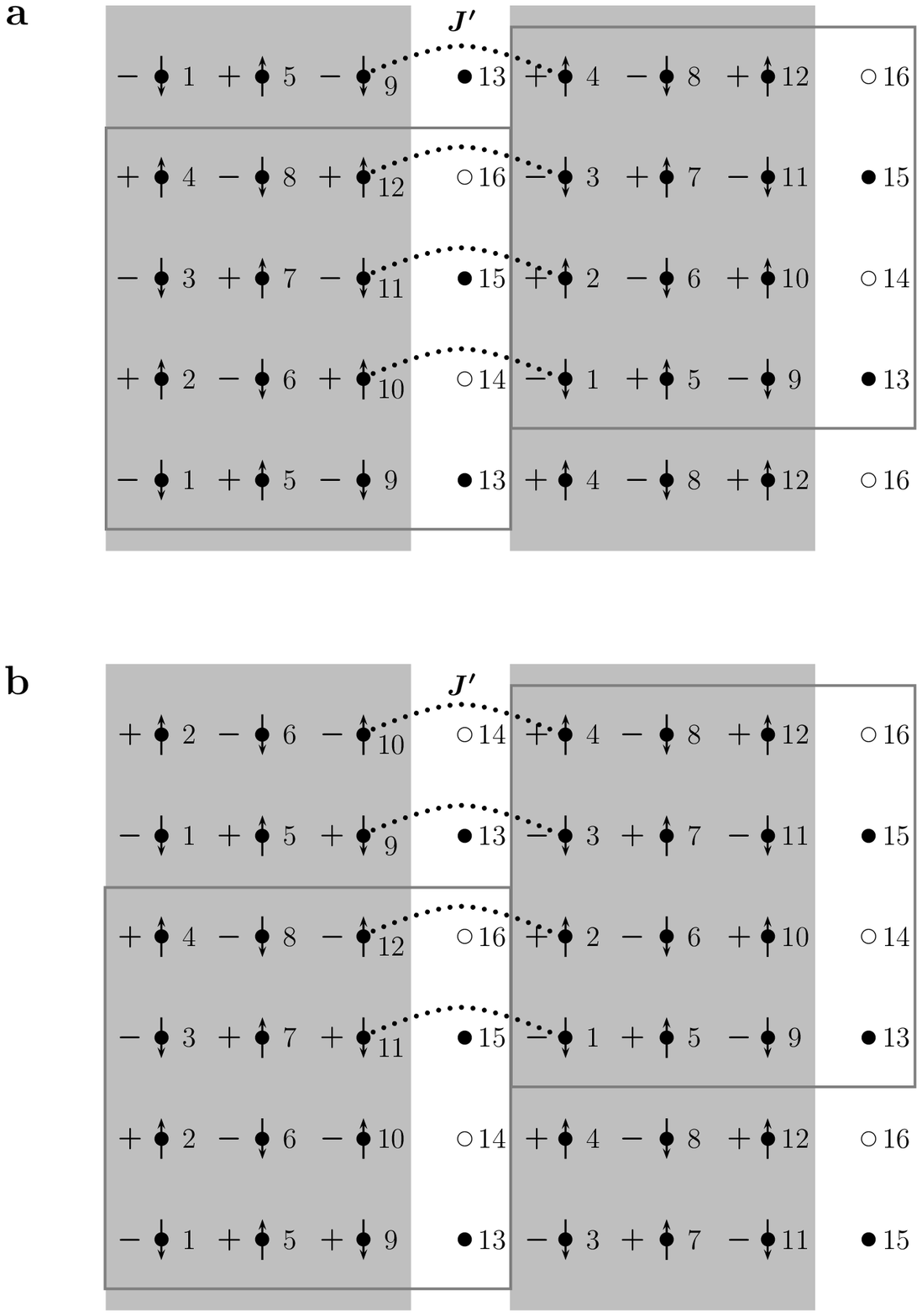}
  \end{center}
  \vspace{-8pt}
  \caption{%
    Finite size geometries with (a) unfrustrated and (b) frustrated
    periodic boundary conditions for site-centered stripe models.  The
    spin stripes are localized by a staggered magnetic field $B$, as
    indicated by the signs in the grey shaded areas.}
  \label{fig:jperp_sc}
\end{figure}

The effective, antiferromagnetic coupling $J'$ between neighboring
spin stripes described by three-leg ladders is induced by the charge
stripes between the ladders, since an antiferromagnetic alignment of
the spins on each side of the charge stripe enhances the mobility of
the holes.  To determine this coupling, we have exactly diagonalized
16 site clusters of itinerant spin 1/2 antiferromagnets described by
the $t$--$J$ model\cite{Zhang-88prb3759, Eskes-88prl1415} with
$J=0.4t$, two holes, and periodic boundary conditions (PBCs), in which
the spin stripes are localized.  The Hamiltonian is given by
\begin{eqnarray}
  \label{eq:htjb}\nonumber
  \hat H_\text{$t$--$J$--$B$} &=&
  -t \sum_{\langle i,j\rangle,\sigma}\hspace{-5pt}
  {\text P}_{\text G} 
  \big(c_{i,\sigma}^{\dagger} c_{j,\sigma}^{\phantom{\dagger}} 
  + \text{h.c.}\big) {\text P}_{\text G}
  + J\sum_{\langle i,j\rangle} \hat{\bs{S}}_i \hat{\bs{S}}_j \\
  &&+ \sum_{i\in \text{shaded area}} B_i\hat S_i^z \, ,
\end{eqnarray}
where the first two sums extend over all nearest-neighbor pairs
$\langle i,j\rangle$, and $B_i=\pm B$ denotes a staggered magnetic
field, as indicated by the signs in Fig.~\ref{fig:jperp_sc}.  The
Gutzwiller projector P$_{\text G}$ eliminates doubly occupied sites.

\begin{table}[t]
  \centering
  \begin{tabular}{|@{\hspace{8pt}}c @{\hspace{8pt}}|
      |@{\hspace{8pt}}c @{\hspace{8pt}}|@{\hspace{8pt}}c 
       @{\hspace{8pt}}|@{\hspace{8pt}}c @{\hspace{8pt}}|}\hline
  \multicolumn{4}{|c|}{\bf Site-centered stripe models} \\\hline
  PBCs & unfrustrated & frustrated & $\Delta$ \\\hline 
  as in & Fig.~\ref{fig:jperp_sc}a & Fig.~\ref{fig:jperp_sc}b 
  & \\\hline\hline
  \multicolumn{4}{|c|}{$t$--$J$ model with spin and charge stripes}\\ 
  \multicolumn{4}{|c|}{$t=2.5$, $J=1$, $B=0.17$, $N=16$, 2 holes} 
  \\\hline
  $E_\text{$t$--$J$--$B$}$ & -21.3409 & -21.2405 & 0.1004 \\
  $E_{B}$ & -0.5415 &&  \\
  $\ev{\hat S^z_1}$ &-0.3140 &&  \\
  $\ev{\hat S^z_5}$ & 0.2411 &&  \\
  $\ev{\hat n_1}$   & 0.9437 &&  \\
  $\ev{\hat n_5}$   & 0.8781 &&  \\\hline\hline
  \multicolumn{4}{|c|}{three-leg Heisenberg ladder with coupling $J'$}\\ 
  \multicolumn{4}{|c|}{$J'=0.071$, $J=1$, $B=0.17$, $N=12$} 
  \\\hline
  $E_\text{$J$--$B$}$ & -8.0068 & -7.9065 & 0.1003 \\\hline
  \end{tabular}
  \caption{Ground state energies and expectation values of the 
    staggered magnetic field term $\hat H_\text{$B$}=\sum B_i\hat S_i^z$,
    of $\hat S_i^z$, and of the electron densities $\hat n_i$ 
    obtained by exactly diagonalization of the $t$--$J$ model 
    \eqref{eq:htjb} as well as the coupled Heisenberg ladders 
    described by \eqref{eq:hjb} for the clusters shown in 
    Fig.~\ref{fig:jperp_sc}a and \ref{fig:jperp_sc}b for unfrustrated 
    and frustrated periodic boundary conditions, respectively.}
  \label{tab:jperp_sc}
\end{table}

We compare the ground state energies we obtain for clusters with
unfrustrated PBCs shown in Fig.~\ref{fig:jperp_sc}a with the ground
state energies we obtain for clusters with frustrated PBCs shown in
Fig.~\ref{fig:jperp_sc}b, in which the 16-site unit cells on the right
are shifted by one lattice spacing to the top.  We then consider
spin-only Heisenberg models of three-leg ladders, which consist of
only the sites in the shaded areas in Figs.~\ref{fig:jperp_sc}a and
\ref{fig:jperp_sc}b, subject to the same staggered field $B_i$, and
couple them antiferromagnetically by $J'$, as indicated.  The
Heisenberg models are described by
\begin{equation}
  \label{eq:hjb}
  \hat H_\text{$J$--$B$} 
  = \sum_{\langle i,j\rangle} J_{ij}\hat{\bs{S}}_i \hat{\bs{S}}_j
  + \sum_{i\in \text{shaded area}} B_i\hat S_i^z \, ,
\end{equation}
where $J_{ij}=J$ for all nearest-neighbor links inside the shaded
areas, but $J_{ij}=J'$ for nearest-neighbor links across the
horizontal boundary lines between those areas.
We again compare the ground state energies for unfrustrated PBCs,
where $J'$ couples sites 10 and 1, 11 and 2, \etc for the three-leg
ladders shown in the shaded areas in Fig.~\ref{fig:jperp_sc}a, with
frustrated PBCs, where $J'$ couples sites 11 and 1, 12 and 2, \etc ,
as shown in Fig.~\ref{fig:jperp_sc}b.  Finally, we determine $J'$ such
that the difference in the ground state energies between frustrated
and unfrustrated PBCs in the $t$--$J$ clusters matches this difference
in the spin-only ladder models.

With $B=0.170J$ we obtain $J'=0.071J$, as detailed in
Tab.~\ref{tab:jperp_sc}.  The value for $B$ is chosen
self-consistently such that the mean value of the staggered
magnetization we obtain for our spin wave theory in Appendix
\ref{sec:staggmag},
\begin{equation}
  \label{eq:staggmag}
  \frac{1}{2}\left(-\ev{\hat s^z_1}+\ev{\hat s^z_2}\right)=0.2903,
\end{equation}
matches the corresponding value in the $t$--$J$ cluster shown in
Fig.~\ref{fig:jperp_sc}a.  For this cluster, however, there are 
two values for the staggered magnetization, depending
on whether we consider the overall magnetization  
\begin{equation}
  \label{eq:staggmag1}
  \frac{1}{2}\left(-\ev{\hat S^z_1}+\ev{\hat S^z_5}\right)=0.2775,  
\end{equation}
or the magnetization on only those sites which are occupied by
electrons (which differs since the holes are not strictly
localized on the chain in between the shaded areas in
Fig.~\ref{fig:jperp_sc}b):
\begin{equation}
  \label{eq:staggmag2}
  \frac{1}{2}\left(-\frac{\ev{\hat S^z_1}}{\ev{\hat n_1}}
    +\frac{\ev{\hat S^z_5}}{\ev{\hat n_5}}\right)=0.3037.
\end{equation}
We assert that these two values constitute lower and upper bounds of
what we would expect in a spin-only model, and chose $B$ such that the
spin wave theory gives the mean value of these bounds (this value is 0.2906).

\begin{figure}[t]
  \begin{center}
    \includegraphics[width=0.90\columnwidth]{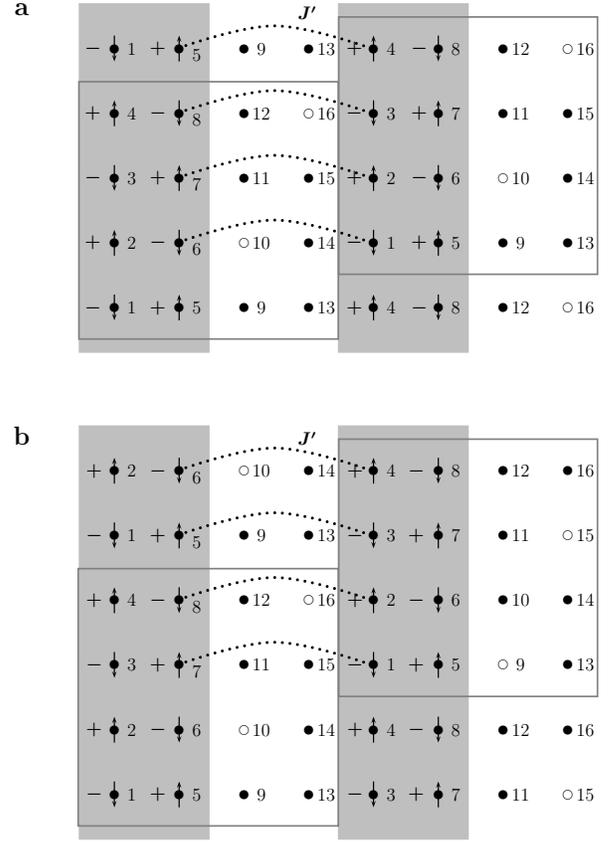}
  \end{center}
  \vspace{-8pt}
  \caption{%
    Finite size geometries with (a) unfrustrated and (b) frustrated
    periodic boundary conditions for bond-centered stripe models.  The
    spin stripes are localized by a staggered magnetic field $B$ as
    indicated by the signs in the grey shaded areas.}
  \label{fig:jperp_bc}
\end{figure}

\begin{table}[t]
  \centering
  \begin{tabular}{|@{\hspace{8pt}}c @{\hspace{8pt}}|
      |@{\hspace{8pt}}c @{\hspace{8pt}}|@{\hspace{8pt}}c 
       @{\hspace{8pt}}|@{\hspace{8pt}}c @{\hspace{8pt}}|}\hline
  \multicolumn{4}{|c|}{\bf Bond-centered stripe models} \\\hline
  PBCs & unfrustrated & frustrated & $\Delta$ \\\hline 
  as in & Fig.~\ref{fig:jperp_bc}a & Fig.~\ref{fig:jperp_bc}b 
  & \\\hline\hline
  \multicolumn{4}{|c|}{$t$--$J$ model with spin and charge stripes}\\ 
  \multicolumn{4}{|c|}{$t=2.5$, $J=1$, $B=0.225$, $N=16$, 2 holes} 
  \\\hline
  $E_\text{$t$--$J$--$B$}$ & -21.3428 & -21.2526 & 0.0902\\
  $E_{B}$ & -0.5395 &&  \\\hline\hline
  \multicolumn{4}{|c|}{two-leg Heisenberg ladder with coupling $J'$}\\ 
  \multicolumn{4}{|c|}{$J'=-0.051$, $J=1$, $B=0.225$, $N=8$} 
  \\\hline
  $E_\text{$J$--$B$}$ & -5.1557 & -5.0655 & 0.0902 \\\hline
  \end{tabular}
  \caption{Ground state energies and the expectation value of the 
    staggered magnetic field term $\hat H_\text{$B$}=\sum B_i\hat S_i^z$
    obtained by exactly diagonalization of
    the $t$--$J$ model \eqref{eq:htjb} as well as the coupled 
    Heisenberg ladders described by \eqref{eq:hjb} for the 
    clusters shown in Fig.~\ref{fig:jperp_bc}a and 
    \ref{fig:jperp_bc}b for unfrustrated and frustrated periodic 
    boundary conditions, respectively.}
  \label{tab:jperp_bc}
\end{table}

For completeness, we also provide the details of the corresponding
calculation for the effective coupling $J'$ between bond-centered
stripes modelled by two-leg ladders, in Tab~\ref{tab:jperp_bc}.  The
calculation 
differs only in that we now use the finite size clusters depicted in
Figs.~\ref{fig:jperp_bc}a and \ref{fig:jperp_bc}b.  The value of the
staggered magnetic field $B=0.225$ is chosen such that the staggered
magnetic field energy $E_B$ described by the last term in
\eqref{eq:htjb} is equal to the value we obtained for the 
three-leg ladder model
listed in see Tab.~\ref{tab:jperp_sc}.  We obtain a ferromagnetic
coupling $J'=-0.051J$.  This value is not large enough to close the
spin gap $\Delta\approx J/2$ of the the individual two-leg ladders,
and hence precludes a description of bond-centered stripes in terms of
spin-only models of coupled two-leg ladders\cite{greiter-10prb144509}.

It should be born in mind that the values we obtain here are only
estimates, as it is impossible to calculate a precise value for a
coupling between spin-only models of stripes, as these constitute a
rather crude approximation themselves.  We are confident, however,
that the coupling for three-leg ladders modelling site-centered stripes
is between $J'=0.05J$ and $J'=0.1J$, and that the absolute value of
the coupling between two-leg ladders describing 
bond-centered stripes is significantly smaller than this coupling.


\section{Analysis of $\bs{\tilde E_{0}}$, $\bs{\hat H_{a0}}$,
  $\bs{\hat H_{c0}}$, $\bs{\hat H_{a1}}$, and $\bs{\hat H_{c2}}$}
\label{sec:other}

In this appendix, we expand and analyze the terms in the Hamiltonian
\eqref{eq:h2d-decompose} which have no influence on the low energy
spectrum in this appendix.  These terms, however, are required for
evaluations of the ground state energy.

\subsection{The bare ground state energy $\bs{\tilde E_{0}}$ }

The bare ground state energy $\tilde E_{0}$ accounts for the constant
terms in the Hamiltonian \eqref{eq:h2d}.  It is given by $\tilde
E_0=\sum_{\bk} 2\tilde\varepsilon_0=N\tilde\varepsilon_0$
with
\begin{eqnarray}\nonumber
  \label{eq:epsilon0tilde}
  \tilde\varepsilon_0 
  &=& \frac{1}{2}-\frac{3}{2} 
  u^2 + J\sum_{\alpha=1}^3 s^z_{\alpha,11} S^z_{\alpha,11}
  +J'\,s^z_{3,11} S^z_{1,11}\\
  &=& -1.45842.
\end{eqnarray}

\subsection{Evaluation of the spectrum of $\bs{\hat H_{a0}}$}
\begin{figure}[t]
    \begin{center}
      \vspace{2pt}
      \includegraphics[height=65mm]{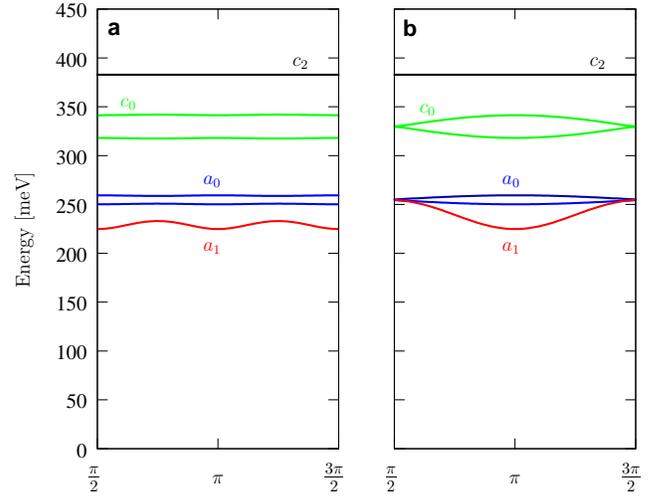}
    \end{center}
      \vspace{-10pt}
  \caption{(Color online)\  Modes described by $\hat H_{a0}$ (blue),
    $\hat H_{c0}$ (green), $\hat H_{a1}$ (red), $\hat H_{c-2}$ (black) 
    plotted as cuts (a) along $(k_x,\pi)$ and (b) along $(\pi,k_y)$ 
    using $J_{\text{exp}}=140\,\text{meV}$.}
  \label{fig:other}
\end{figure}

The term $\hat H_{a0}$ is given by
\begin{eqnarray}\nonumber
  \label{eq:ha0}
  \hat H_{a0}&=&\sum_{\bk}\Big[
  \varepsilon_{a0} \big(a^{\dagger}_{0,\bk} a^{\phantom{\dagger}}_{0,\bk} 
  +A^{\dagger}_{0,\bk} A^{\phantom{\dagger}}_{0,\bk} \big)\\
  &&\hspace{15pt}+\;\xi_{a0,\bk} \big(a^{\dagger}_{0,\bk} 
  + a^{\phantom{\dagger}}_{0,\bk}\big)
  \big(A^{\dagger}_{0,\bk} + A^{\phantom{\dagger}}_{0,\bk} 
  \big)\Big],\hspace{15pt}
\end{eqnarray}
with 
\begin{eqnarray} \label{eq:epsilon_a0}\nonumber
\varepsilon_{a0}\!&=&\!
-\frac{1}{2}+\frac{3}{2}u^2
-2J\sum_{\alpha=1}^3s^z_{\alpha,11}S^z_{\alpha,11} 
-2J' s^z_{3,11}S^z_{1,11},\\[-4pt] \\[-4pt]\label{eq:xi_a0k} \nonumber 
\xi_{a0,\bk}\!&=&\!
J'\cos(4k_x)\, s_{3,13}^z S_{1,13}^z 
+J\cos k_y \sum_{\alpha=1}^3 s_{\alpha,13}^z S_{\alpha,13}^z.\\[-4pt]
\end{eqnarray}
To a reasonable approximation, we obtain low energy modes
described by $\hat H_{a0}$ by diagonalizing \eqref{eq:ha0} at each
point in $\bk$ space in the reduced Hilbert space spanned by
\begin{displaymath}
  \sket{\tilde 0},\ a^{\dagger}_{0}\sket{\tilde 0},\ 
  A^{\dagger}_{0}\sket{\tilde 0},\ \text{and}\ 
  a^{\dagger}_{0} A^{\dagger}_{0}\sket{\tilde 0},
\end{displaymath}
where 
\begin{equation}
  \label{eq:vac_tilde}
  \sket{\tilde 0}
  \equiv\prod_{i\in {\cal A}} \sket{\tilde{b}_{-1/2}}_i^{\phantom{\dagger}}
  \cdot\prod_{j\in {\cal B}} \sket{\tilde{b}_{1/2}}_j^{\phantom{\dagger}}
\end{equation}
is the bare vacuum unrenormalized by spin wave theory.  This yields
two almost dispersionless modes
\begin{equation}
  \label{eq:a0_modes}
  (\omega_{a0,\bk})_{1/2}^{\phantom{\dagger}}
  =\sqrt{\varepsilon_{a0}^2+\xi_{a0,\bk}^2}\,\pm\,\xi_{a0,\bk},
\end{equation}
with energies of about $\varepsilon_{a0}=2.07$, or
$290\,\text{meV}$ if we assume
$J_{\text{exp.}}=140\,\text{meV}$.  Cuts of the dispersions of these
two modes are shown in blue (color online) in Fig.~\ref{fig:other}.  
$\hat H_{a0}$ also gives rise to a contribution 
\begin{equation}
  \label{eq:a0_gsenergy}
  E_{a0}=\sum_{\bk} 
  \Big(\varepsilon_{a0}-\sqrt{\varepsilon_{a0}^2+\xi_{a0,\bk}^2}\,\Big)
  = -0.00008\,N 
\end{equation}
to the ground state energy.  (Here $N$ denotes the number of rungs,
which implies that the sum extends over $\frac{N}{2}$ values for $\bk$.)

\subsection{Evaluation of the spectrum of $\bs{\hat  H_{c0}}$}
A similar analysis of
\begin{eqnarray}\nonumber
  \label{eq:hc0}
  \hat H_{c0}&=&\sum_{\bk}\Big[
  \varepsilon_{c0} \big(c^{\dagger}_{0,\bk}c^{\phantom{\dagger}}_{0,\bk} 
  +C^{\dagger}_{0,\bk}C^{\phantom{\dagger}}_{0,\bk} \big)\\
  &&\hspace{15pt}
  +\;\xi_{c0,\bk} \big(c^{\dagger}_{0,\bk} + c^{\phantom{\dagger}}_{0,\bk}\big)
 \big(C^{\dagger}_{0,\bk} + C^{\phantom{\dagger}}_{0,\bk} \big)\Big],\hspace{15pt}
\end{eqnarray}
with 
\begin{eqnarray}\label{eq:epsilon_c0}\nonumber
\varepsilon_{c0}&=&
\frac{3}{2}\left(u^2-v^2\right)+ 
2J\sum_{\alpha=1}^3s^z_{\alpha,11}(S^z_{\alpha,66}-S^z_{\alpha,11})\hspace{15pt}\\
&+& 2J'\, s^z_{3,11}(S^z_{1,66}-S^z_{1,11}),\\\label{eq:xi_c0k}\nonumber 
\xi_{c0,\bk}&=&
J'\cos(4k_x)\, s_{3,16}^z S_{1,16}^z
+J\cos k_y \sum_{\alpha=1}^3 s_{\alpha,16}^z S_{\alpha,16}^z, \\[-4pt]
\end{eqnarray}
yields two additional, almost dispersionless modes 
\begin{equation}
  \label{eq:c0_modes}
  (\omega_{c0,\bk})_{1/2}^{\phantom{\dagger}}
  =\sqrt{\varepsilon_{c0}^2+\xi_{c0,\bk}^2}\,\pm\,\xi_{c0,\bk},
\end{equation}
with energies of about $\varepsilon_{c0}=2.71$, or $380\,\text{meV}$,
which is shown in green (color online) in Fig.~\ref{fig:other}.  $\hat
H_{c0}$ also gives rise to a contribution
\begin{equation}
  \label{eq:c0_gsenergy}
  E_{c0}=\sum_{\bk} 
  \Big(\varepsilon_{c0}-\sqrt{\varepsilon_{c0}^2+\xi_{c0,\bk}^2}\,\Big)
  = -0.00048\,N 
\end{equation}
to the ground state energy.

\subsection{Evaluation of the spectrum of $\bs{\hat H_{a1}}$}

The term
\begin{eqnarray}\nonumber
  \label{eq:ha1}
  \hat H_{a1}&=&\sum_{\bk}\Big[
  \varepsilon_{a1} \big(a^{\dagger}_{1,\bk}a^{\phantom{\dagger}}_{1,\bk} 
  +A^{\dagger}_{-1,\bk}A^{\phantom{\dagger}}_{-1,\bk} \big)\\
  &&\hspace{15pt}+\;\xi_{a1,\bk} \big(
  a^{\dagger}_{1,\bk} A^{\dagger}_{-1,\bk}
  + a^{\phantom{\dagger}}_{1,\bk} A^{\phantom{\dagger}}_{-1,\bk}
  \big)\Big],\hspace{15pt}
\end{eqnarray}
with $\varepsilon_{a1}=\varepsilon_{a0}$ as given in
\eqref{eq:epsilon_a0} and
\begin{eqnarray}\nonumber 
\xi_{a1,\bk}\!&=&\!J'\cos(4k_x) s^+_{3,41}S^-_{1,41}
+J\cos k_y\,\sum_{\alpha=1}^3 s^+_{\alpha,41} S^-_{\alpha,41},\\[-4pt]
\end{eqnarray}
can be diagonalized by a Bogoliubov transformation.  We obtain
\begin{equation}
  \label{eq:ha1bogo}
  \hat H_{a1}=\sum_{\bk} \big[\omega_{a1,\bk} 
  \big(\alpha_{1,\bk}^{\dagger}\alpha_{1,\bk}^{\phantom{\dagger}}+
       \alpha_{2,\bk}^{\dagger}\alpha_{2,\bk}^{\phantom{\dagger}}\big)
       + \omega_{a1,\bk} 
       - \varepsilon_{a1}\Big]
\end{equation}
with 
\begin{equation}
  \label{eq:a1_modes}
  \omega_{a1,\bk}=\sqrt{\varepsilon_{a1}^2-\xi_{a1,\bk}^2}.
\end{equation}
It yields a two-fold degenerate, weakly dispersing mode with an energy
of about $1.95$, or $273\,\text{meV}$, which is shown in red (color
online) in Fig.~\ref{fig:other}, as well as a contribution
\begin{equation}
  \label{eq:a1_gsenergy}
  E_{a1}=\sum_{\bk}\Big(\omega_{a1,\bk}- \varepsilon_{a1}\Big) 
  =-0.05363\,N 
\end{equation}
to the ground state energy.

\subsection{Evaluation of the spectrum of $\bs{\hat H_{c2}}$}

Finally, 
\begin{eqnarray}
  \label{eq:hc2}
  \hat H_{c2} \,&=&\,\sum_{\bk}
  \omega_{c2} \big(c^{\dagger}_{2,\bk}c^{\phantom{\dagger}}_{2,\bk} 
  +C^{\dagger}_{-2,\bk}C^{\phantom{\dagger}}_{-2,\bk} \big)
\end{eqnarray}
with 
\begin{eqnarray}\nonumber
\omega_{c2}&=&
\frac{3}{2}u^2 
+ 2J\sum_{\alpha=1}^3s^z_{\alpha,11}(S^z_{\alpha,88}-S^z_{\alpha,11})  \\ 
&+& 2J'\, s^z_{3,11}(S^z_{1,88}-S^z_{1,11}),
\end{eqnarray}
describes a two-fold degenerate, completely dispersionless mode with an
energy of $\omega_{c2}=3.12$,
or $436\,\text{meV}$.

\section{Ground state energy}
\label{sec:groundstate}

To evaluate the ground state energy $E_0$, we collect the
contributions from \eqref{eq:epsilon0tilde}, \eqref{eq:a0_gsenergy},
\eqref{eq:c0_gsenergy}, \eqref{eq:a1_gsenergy}, and
\eqref{eq::b1c1c-1_gsenergy}.  This yields
\begin{equation}
  \label{eq:gsenergy}
  E_0=-1.73378\,N,
\end{equation}
where $N$ is the number of rungs.  This number is in good
agreement with what we would expect from the results of exact
diagonalizations of 
\begin{equation}
  \label{eq:hj}
  \hat H_\text{$J$} 
  = \sum_{\langle i,j\rangle} J_{ij}\hat{\bs{S}}_i \hat{\bs{S}}_j
\end{equation}
for small clusters of 12, 18 and 24 sites with unfrustrated boundary
conditions, as shown for $N=4$ rungs by the shaded areas in
Fig.~\ref{fig:jperp_sc}a.  Specifically, we obtain $E_0 = -1.827\,N$
for a cluster with $N=8$ rungs, $J'=0.07$, and unfrustrated boundary
conditions with a shift of 3 lattice spacings. If we compare the
nearest-neighbor spin-spin correlation we obtain from exactly
diagonalizing the same 24 site cluster with $J'=1$,
$\ev{\hat{\bs{S}}_i\hat{\bs{S}}_j}=-0.343$, to the the value predicted
by standard two-dimensional linear spin wave
theory\cite{anderson52pr694},
$\ev{\hat{\bs{S}}_i\hat{\bs{S}}_j}=-0.329$, we are led to estimate
that the bond operator spin wave theory developed here should yield a
number around
\begin{equation}
  \label{eq:gsenergyest}
  E_0 = -1.827\,N \cdot\frac{-0.329}{-0.3432} =-1.751\,N.
\end{equation}
This differs only by 1\% from \eqref{eq:gsenergy}.






\section{Matrix Elements}
\label{sec:matrixelements}

The dynamical structure factor measured in neutron scattering is
given by 
\begin{equation}
  \label{eq:chi+-}
\chi^{+-}(\bs{k},\omega) = 
\sum_{n}|\bra{0} \hat{S}^+_{\bs{k}} \ket{n}|^2\delta(\omega-\omega_n),
\end{equation}
where $\ket{0}$ is the ground state and the sum extends over all 
excited states $\ket{n}$ with energy $\omega_n$, and 
\begin{equation}
  \label{eq:s+k}
  \hat{S}^{+}_{\bs{k}}=\sum_{l}e^{-i\bs{k}\bs{r}_{l}}\hat{S}_{l}^+
\end{equation}
is the Fourier transform of the spin raising operator $\hat{S}_{l}$
at lattice site $l$ with respect to original lattice, \ie the sum
runs over all lattice sites. This implies 
\begin{eqnarray}
  \label{eq:rofR}
  \bs{r}_{i,\alpha}&=&\bs{R}_i + \begin{pmatrix}\alpha-2\\0\end{pmatrix},
  \\[5pt]
  \bs{r}_{j,\alpha}&=&\bs{R}_j + \begin{pmatrix}\alpha-2\\0\end{pmatrix},
\end{eqnarray}
with $\bs{R}_i$ and $\bs{R}_j$ as indicated in Fig.~\ref{fig:2dm} for
sublattices $A$ and $B$, respectively.  In analogy to Fourier
transforms of the bosonic creation and annihilation operators
\eqref{eq:fta} and \eqref{eq:ftb}, we further Fourier transforms of
the spin operators with respect to the rung sublattices $\cal A$ and
$\cal B$ according to
\begin{eqnarray}
  \label{eq:s+kalpha}
  \hat{s}^+_{\bs{k},\alpha} &=& 
  \sqrt{\frac{2}{N}}\sum_{i\in {\cal A}}e^{-i\bs{k}\bs{R}_i}\hat{s}^+_{i,\alpha}
  \\
  \hat{S}^+_{\bs{k},\alpha} &=&
  \sqrt{\frac{2}{N}}\sum_{j\in {\cal B}}e^{-i\bs{k}\bs{R}_j}\hat{S}^+_{j,\alpha},
\end{eqnarray}
and express the operator \eqref{eq:s+k} in terms of them:
\begin{eqnarray}
  \label{eq:s+ks+kalpha}\nonumber
  \hat{S}^{+}_{\bs{k}}
  \!&=&\! \sqrt{\frac{2}{N}}\biggl\{\,
    \sum_{i\in{\cal A}}\sum_\alpha 
    e^{-i\bs{k}\bs{r}_{i,\alpha}}\hat{s}_{i,\alpha}^+ +
    \sum_{j\in{\cal B}}\sum_\alpha 
    e^{-i\bs{k}\bs{r}_{j,\alpha}}\hat{S}_{j,\alpha}^+ 
  \biggr\}\\ \nonumber
  \!&=&\! \sum_{\alpha}e^{-ik_x(\alpha-2)}s^{+}_{\bs{k},\alpha} + 
  \sum_{\alpha}e^{-ik_x(\alpha-2)}S^{+}_{\bs{k},\alpha}\\
  \!&=&\! \sum_{\alpha}e^{-ik_x(\alpha-2)}
  \bigl(\hat{s}^+_{\bs{k},\alpha}+\hat{S}^+_{\bs{k},\alpha} \bigr)
\end{eqnarray}

\begin{figure}[t]
    \begin{center}
      \psfrag{c}{{$0$}}
      \psfrag{d}{$\pi$}
      \psfrag{e}{{$2\pi$}}
      \includegraphics[height=55mm]{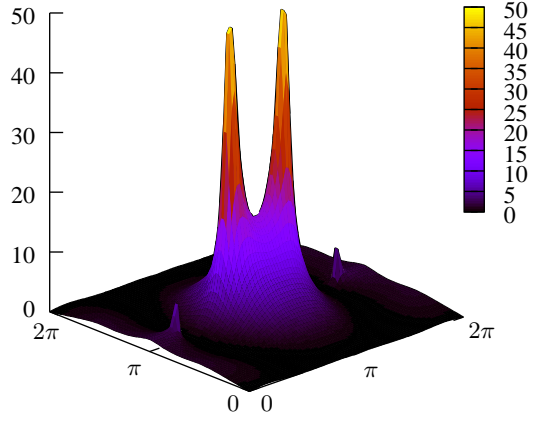}
    \end{center}
    \vspace{-10pt}
    \caption{(Color online) Numerical evaluation of the matrix elements 
      $|\sbra{0}\hat{S}^{+}_{\bs{k}}\sket{\gamma_{1,\bs{k}}}|^2$ for
      the entire Brillouin zone $\bs{k}\in [0,2\pi]\times [0,2\pi]$. 
      Note the strong enhancement aground the antiferromagnetic
      ordering wave vectors $\bs{k}=(\pi\pm\pi/4,\pi)$.}
  \label{fig:matrixelements3d}
\end{figure}
As we are interested only in the contribution of the low-energy mode
$\omega_{\bk,1}$ to $\chi^{+-}(\bs{k},\omega)$, the only matrix 
element we need to evaluate is
\begin{eqnarray}\label{ma_el}
  \bigl|\sbra{0}\hat{S}^{+}_{\bs{k}}\sket{\gamma_{1,\bs{k}}}\bigr|^2 = 
  \bigl|\sbra{0}\hat{S}^{+}_{\bs{k}}\gamma^{\dagger}_{1,\bs{k}}\sket{0}\bigr|^2
\end{eqnarray}
for all values of $\bs{k}$.  (The second low-energy mode
$\omega_{\bk,2}$, which is degenerate with the first, does not
contribute to \eqref{ma_el} and hence to $\chi^{+-}(\bs{k},\omega)$,
but instead yields a contribution to $\chi^{-+}(\bs{k},\omega)$ which
is identical to the one we calculate below.)  Keeping only terms which
contribute to this mode and are linear in the expansion \eqref{eq:sa}
of $\hat{s}_{\alpha}^{+}$, we obtain for sublattice $\cal A$ 
\begin{equation}
  \label{eq:sa+alpha}\nonumber
  \hat{s}_{\alpha}^{+} = 
  s_{\alpha,21}^{+}b_1^{\dagger} +
  s_{\alpha,71}^{+}c_1^{\dagger} + 
  s_{\alpha,15}^{+}c_{-1}
\end{equation}\\
or in Fourier space with \eqref{eq:s+kalpha}, \eqref{eq:fta}, and 
\eqref{eq:psi_op}
\begin{eqnarray}
  \label{eq:sa+kalpha}\nonumber
  \hat{s}_{\bs{k},\alpha}^{+} &=& 
  s_{\alpha,21}^{+}b_{1,\bs{k}}^{\dagger} +
  s_{\alpha,71}^{+}c_{1,\bs{k}}^{\dagger} + 
  s_{\alpha,15}^{+}c_{-1,-\bs{k}}^{\phantom{\dagger}}\\
  &=&\bigl(0,s^{+}_{\alpha,21},0,s^{+}_{\alpha,71},0,s^{+}_{\alpha,15}\bigr)
  \cdot\Psi_{\bs{k}}
\end{eqnarray}
and similarly with \eqref{eq:sb} and \eqref{eq:ftb} for $\cal B$ 
\begin{eqnarray}
  \label{eq:sb+kalpha}\nonumber
  \hat{S}_{\bs{k},\alpha}^{+} &=& 
  S_{\alpha,12}^{+}B_{-1,\bs{k}}^{\phantom{\dagger}} +
  S_{\alpha,17}^{+}C_{-1,\bs{k}}^{\phantom{\dagger}} +  
  S_{\alpha,51}^{+}C_{1,-\bs{k}}^{\dagger}\\
  &=&\bigl(S^{+}_{\alpha,12},0,S^{+}_{\alpha,17},0,S^{+}_{\alpha,51},0\bigr)
  \cdot\Psi_{\bs{k}}.
\end{eqnarray}
We then use $\Psi_{\bs{k}} = M_{\bs{k}}\Gamma_{\bs{k}}$ and
\eqref{eq:gamma_ops} to express $\hat{s}_{\bs{k},\alpha}^{+}$ and
$\hat{S}_{\bs{k},\alpha}^{+}$ in terms of $\gamma_{i,\bs{k}}$ and
$\gamma_{i,\bs{k}}^{\dagger}$, recall
$S^+_{\alpha,ji}=s^+_{\alpha,ij}$, and finally obtain
\begin{eqnarray}
  \label{eq:matelem}\nonumber
  \bigl|\sbra{0}\hat{S}^{+}_{\bs{k}}\sket{\gamma_{1,\bs{k}}}\bigr|^2 =
  \Bigl|\Bigr.\sum_{\alpha=1}^3 e^{-ik_x(\alpha-2)}\bigl\{\bigr.
    s^+_{\alpha,21}(M_{\bs{k},11}\!\!&\!+\!&\!\!M_{\bs{k},21}) \\\nonumber 
    +\,s^+_{\alpha,71}(M_{\bs{k},31}\!+\!M_{\bs{k},41}) + 
    s^+_{\alpha,15}(M_{\bs{k},51}\!\!&\!+\!&\!\!M_{\bs{k},61})
  \bigl.\bigr\}\Bigl.\Bigr|^2\\
\end{eqnarray}
Note that $\chi^{zz}(\bs{k},\omega)=0$ as there is no term linear in 
$b_1^{\dagger}$, $c_1^{\dagger}$, or $c_{-1}$ in the expansion \eqref{eq:sa}
for $\hat{s}_{\alpha}^{z}$.\\[5pt]

\section{Staggered magnetization}
\label{sec:staggmag}

The staggered magnetizations on the outer and inner chains of our
three-leg ladders,
$\ev{\hat{s}^z_1}=\ev{\hat{s}^z_3}=-\ev{\hat{S}^z_1}=-\ev{\hat{S}^z_3}$
and $\ev{\hat{s}^z_2}=-\ev{\hat{S}^z_2}$, respectively, are given in
by the the bare values 
$s_{\alpha,11}^z=-S_{\alpha,11}^z$ with $\alpha=1,2,3$ minus
corrections from the individual terms in the Hamiltonian
\eqref{eq:h2d-decompose}.  From 
\eqref{eq:sz1}--\eqref{eq:sz3} 
with \eqref{eq:phi}, we obtain for the bare values
\begin{eqnarray}
\label{eq:sz1_bare}
\langle \hat{s}_1^z\rangle_{\mathrm{bare}}\! &=&\! s_{1,11}^z = -0.4633\\
\label{eq:sz2_bare}
\langle \hat{s}_2^z\rangle_{\mathrm{bare}}\! &=&\! s_{2,11}^z = 0.4265
\end{eqnarray}
We expect that the largest corrections arise from
$\hat{H}_{b1,c1,c-1}$, as this part contains the only low energy mode
of the theory.  The for this part relevant terms in the expansion
\eqref{eq:sa} for $\hat{s}^z_{\alpha}$ are
\begin{multline}
 b_1^{\dagger}b_1^{\phantom{\dagger}}\left(s^z_{\alpha,22}-s^z_{\alpha,11}\right) +
c_1^{\dagger}c_1^{\phantom{\dagger}}\left(s^z_{\alpha,77}-s^z_{\alpha,11}\right)
\\\nonumber
+c_{-1}^{\dagger}c_{-1}^{\phantom{\dagger}}
\left(s^z_{\alpha,55}-s^z_{\alpha,11}\right) +  
\left(b_1^{\dagger}c_1^{\phantom{\dagger}}
+c_1^{\dagger}b_1^{\phantom{\dagger}}\right) s^z_{\alpha,27}\,.
\end{multline}
For convenience, we define
\begin{eqnarray}\nonumber
\rho_{\alpha,1} &=& \left(s^z_{\alpha,22}-s^z_{\alpha,11}\right)\\
\rho_{\alpha,2} &=& \left(s^z_{\alpha,77}-s^z_{\alpha,11}\right)\\\nonumber
\rho_{\alpha,3} &=& \left(s^z_{\alpha,55}-s^z_{\alpha,11}\right)\, .
\end{eqnarray}
and express the operators $b_1,c_1,c_{-1}^\dagger$ as well as their
hermitian conjugates via $\Psi = M \Gamma$ through the $\gamma$
operators (see \eqref{eq:psi_opdag}), \eqref{eq:defofm}, and
\eqref{eq:gamma_ops} above).  Using
$\gamma^{\phantom{\dagger}}_{j}|0\rangle=\langle0|\gamma^{\dagger}_{j}=0$
and $\langle 0|\gamma_i\gamma_j^{\dagger}|0\rangle =\delta_{ij}$, we
obtain for the corrections from $\hat{H}_{b1,c1,c-1}$
\begin{widetext}
\begin{eqnarray}\nonumber
\langle \hat{s}_{\alpha}^z\rangle_{\hat{H}_{b1,c1,c-1}} 
\!=&&\hspace{-10pt} \frac{2}{N}\sum_{\bs{k}}
\bigg\{
\Big[
\rho_{\alpha,1} M_{\bs{k},21}M^T_{\bs{k},12} +
\rho_{\alpha,2} M_{\bs{k},41}M^T_{\bs{k},14} +
\rho_{\alpha,3} M_{\bs{k},62}M^T_{\bs{k},26} + 
s^z_{\alpha,27}
\left(M_{\bs{k},21}M^T_{\bs{k},14} + M_{\bs{k},41}M^T_{\bs{k},12}\right)
\Big]
\\\nonumber  &&\hspace{18pt}+
\Big[
\rho_{\alpha,1}M_{\bs{k},23}M^T_{\bs{k},32} +
\rho_{\alpha,2} M_{\bs{k},43}M^T_{\bs{k},34} + 
\rho_{\alpha,3}M_{\bs{k},64}M^T_{\bs{k},46} + 
s^z_{\alpha,27}
\left(M_{\bs{k},23}M^T_{\bs{k},34} + M_{\bs{k},43}M^T_{\bs{k},32} \right)
\Big]
\\\nonumber  &&\hspace{18pt}+
\Big[
\rho_{\alpha,1}M_{\bs{k},26}M^T_{\bs{k},62} + 
\rho_{\alpha,2}M_{\bs{k},46}M^T_{\bs{k},64} + 
\rho_{\alpha,3}M_{\bs{k},65}M^T_{\bs{k},56} +
s^z_{\alpha,27}
\left( M_{\bs{k},26}M^T_{\bs{k},64} + M_{\bs{k},46}M^T_{\bs{k},62}\right)
\Big]
\bigg\}. 
\label{eq:sz_corr_b1c1c-1}
\end{eqnarray}
\end{widetext}
In this sum, the terms in the first pair of square brackets originate
from the low energy eigenmodes $\gamma_{1},\gamma_{2}$ in the energy
range from 0 to about 190$\,$meV in Fig.~\ref{fig:h_bcc}, the terms in the
second pair originate form $\gamma_{3},\gamma_{4}$ at about 290$\,$meV,
and the terms in the third pair originate from $\gamma_{5},\gamma_{6}$
at about 460$\,$meV.  Evaluation yields 
\begin{eqnarray}\nonumber
\langle \hat{s}_{1}^z\rangle_{\hat{H}_{b1,c1,c-1}} 
&=& 0.1752 + 0.0018 + 0.0003\\ &=& 0.1773   \label{eq:sz1_corr_b1c1c-1}
\\\nonumber
\langle \hat{s}_{2}^z\rangle_{\hat{H}_{b1,c1,c-1}} 
&=& -0.0864 - 0.0210 - 0.0001 \\ &=& -0.1075 \label{eq:sz2_corr_b1c1c-1}.
\end{eqnarray}
for the outer and inner chains, respectively.  As expected, the low energy
mode we compare to the experiment\cite{tranquada-04n534} yields the
dominant contribution.

The corrections arising from $\hat{H}_{a1}$ are evaluated in complete
analogy.  The Bogoliubov transformation $\Psi_a = M_a \Gamma_a$ with
$\Psi_a\equiv (A_{-1},a_1^{\dagger})^T$ is now only two-dimensional,
and the only contribution comes from the term
$-s^z_{\alpha,11}a^{\dagger}_{1}a^{\phantom{\dagger}}_1$ in
\eqref{eq:sa}:
\begin{equation}
\langle \hat{s}_{\alpha}^z\rangle_{\hat{H}_{a1}} 
= -\frac{2}{N}\sum_{\bk} s^z_{\alpha,11}M_{a,\bs{k},21}M_{a,\bs{k},12}^T.
\end{equation}
Evaluation yields 
\begin{eqnarray}
\langle \hat{s}_{1}^z\rangle_{\hat{H}_{a1}} &=& 0.0127,
\label{eq:sz1_corr_a1}\\
\langle \hat{s}_{2}^z\rangle_{\hat{H}_{a1}} &=& -0.0117.
\label{eq:sz2_corr_a1}
\end{eqnarray}

Finally, $\hat{H}_{a0}$ and $\hat{H}_{c0}$ give rise to corrections
\begin{eqnarray}
\langle \hat{s}_{\alpha}^z\rangle_{\hat{H}_{a0}} &=& 
\left(-s_{\alpha,11}^z\right)\frac{2}{N}\sum_{\bk}\frac{1}{N^2_{a0,\bk}},
\end{eqnarray}
\begin{eqnarray}
\langle \hat{s}_{\alpha}^z\rangle_{\hat{H}_{c0}} &=& 
\left(s_{\alpha,66}-s_{\alpha,11}^z\right)
\frac{2}{N}\sum_{\bk}\frac{1}{N^2_{c0,\bk}}\quad
\end{eqnarray}
where
\vspace{-20pt}
\begin{eqnarray}
N^2_{ao,\bk} &=& \left(\frac{\epsilon_{a0} 
+ \sqrt{\epsilon_{a0}^2+\xi_{a0,\bk}^2 }}{\xi_{a0,\bk}}\right)^2 +1,\\
N^2_{ao,\bk} &=& \left(\frac{\epsilon_{c0} 
+ \sqrt{\epsilon_{c0}^2+\xi_{c0,\bk}^2 }}{\xi_{c0,\bk}}\right)^2 +1
\\\nonumber
\end{eqnarray}
with $\epsilon_{a0}$ and $\xi_{a0,\bk}$ as given in
\eqref{eq:epsilon_a0} and \eqref{eq:xi_a0k} and $\epsilon_{c0}$ and
$\xi_{c0,\bk}$ as given in \eqref{eq:epsilon_c0} and
\eqref{eq:xi_c0k}.  Evaluation yields 
\vspace{-5pt}
\begin{eqnarray}
\langle \hat{s}_{1}^z\rangle_{\hat{H}_{a0}} +
\langle \hat{s}_{1}^z\rangle_{\hat{H}_{c0}} &=& 0.0001,
\label{eq:sz1_corr_a0c0}\\
\langle \hat{s}_{2}^z\rangle_{\hat{H}_{a0}} +
\langle \hat{s}_{2}^z\rangle_{\hat{H}_{c0}}  &=& -0.0001,
\label{eq:sz2_corr_a0c0}
\end{eqnarray}
\ie negligible small contributions.

Summing up \eqref{eq:sz1_bare}, \eqref{eq:sz1_corr_b1c1c-1},
\eqref{eq:sz1_corr_a1}, \eqref{eq:sz1_corr_a0c0}, and
\eqref{eq:sz2_bare}, \eqref{eq:sz2_corr_b1c1c-1},
\eqref{eq:sz2_corr_a1}, \eqref{eq:sz2_corr_a0c0}, we obtain
\begin{eqnarray}
  \label{eq:sz1_ev}
  \langle \hat{s}_1^z\rangle &=& -0.2732,\\
  \label{eq:sz2_ev}
  \langle \hat{s}_2^z\rangle &=& 0.3072
\end{eqnarray}
for the staggered magnetizations on the outer and inner chains of the
three-leg ladders, respectively.


\begin{thebibliography}{10}

\bibitem{zaanen-06np138}
J. Zaanen {\it et~al.}, Nature Physics {\bf 2},  138  (2006).

\bibitem{orenstein-00s468}
J. Orenstein and A.~J. Millis, Science {\bf 288},  468  (2000).

\bibitem{Zhang-88prb3759}
F.~C. Zhang and T.~M. Rice, Phys. Rev. B {\bf 37},  3759  (1988).

\bibitem{Eskes-88prl1415}
H. Eskes and G.~A. Sawatzky, Phys. Rev. Lett. {\bf 61},  1415  (1988).

\bibitem{fong-95prl316}
H.~F. Fong, B. Keimer, P.~W. Anderson, D. Reznik, F. Do\ifmmode~\breve{g}\else
  \u{g}\fi{}an, and I.~A. Aksay, Phys. Rev. Lett. {\bf 75},  316  (1995).

\bibitem{bourges-00s1234}
P. Bourges, Y. Sidis, H.~F. Fong, L.~P. Regnault, J. Bossy, A. Ivanov, and B.
  Keimer, Science {\bf 288},  1234  (2000).

\bibitem{zaanen-89prb7391}
J. Zaanen and O. Gunnarsson, Phys. Rev. B {\bf 40},  7391  (1989).

\bibitem{kato-90jpsj1047}
M. Kato, K. Machida, H. Nakanishi, and M. Fujita, J. Phys. Soc. Jpn. {\bf 59},
  1047  (1990).

\bibitem{tranquada-95n561}
J.~M. Tranquada, B.~J. Sternlieb, J.~D. Axe, Y. Nakamura, and S. Uchida, Nature
  {\bf 375},  561  (1995).

\bibitem{emery-99pnas8814}
V.~J. Emery, S. Kivelson, and J. Tranquada, Proc. Natl. Acad. Sci. U.S.A. {\bf
  96},  8814  (1999).

\bibitem{zaanen-01pmb1485}
J. Zaanen, O.~Y. Osman, H.~V. Kruis, Z. Nussinov, and J. Tworzyd\l{}o, Phil.
  Mag. B {\bf 81},  1485  (2001).

\bibitem{mook-02prl097004}
H.~A. Mook, P. Dai, and F. Do\ifmmode~\breve{g}\else \u{g}\fi{}an, Phys. Rev.
  Lett. {\bf 88},  097004  (2002).

\bibitem{kivelson-03rmp1201}
S.~A. Kivelson, I.~P. Bindloss, E. Fradkin, V. Oganesyan, J.~M. Tranquada, A.
  Kapitulnik, and C. Howald, Rev. Mod. Phys. {\bf 75},  1201  (2003).

\bibitem{berg-09njp115004}
E. Berg, E. Fradkin, S.~A. Kivelson, and J. Tranquada, New J. Phys. {\bf 11},
  115004  (2009).

\bibitem{tranquada-04n534}
J. Tranquada, H. Woo, T. Perring, H. Goka, G. Gu, G. Xu, M. Fujita, and K.
  Yamada, Nature {\bf 429},  534  (2004).

\bibitem{fauque-07prb214512}
B. Fauqu\'{e}, Y. Sidis, L. Capogna, A. Ivanov, K. Hradil, C. Ulrich, A.~I.
  Rykov, B. Keimer, and P. Bourges, Phys. Rev. B {\bf 76},  214512  (2007).

\bibitem{dagotto-96s618}
E. Dagotto and T.~M. Rice, Science {\bf 271},  618  (1996).

\bibitem{shelton-96prb8521}
D.~G. Shelton, A.~A. Nersesyan, and A.~M. Tsvelik, Phys. Rev. B {\bf 53},  8521
   (1996).

\bibitem{greiter02prb054505}
M. Greiter, Phys. Rev. B {\bf 66},  054505  (2002).

\bibitem{Xu-07prb014508}
G. Xu, J.~M. Tranquada, T.~G. Perring, G.~D. Gu, M. Fujita, and K. Yamada,
  Phys. Rev. B {\bf 76},  014508  (2007).

\bibitem{hinkov-04n650}
V. Hinkov, S. Pailhes, P. Bourges, Y. Sidis, A. Ivanov, A. Kulakov, C.~T. Lin,
  D.~P. Chen, C. Bernhard, and B. Keimer, Nature {\bf 430},  650  (2004).

\bibitem{hinkov-07np780}
V. Hinkov, P. Bourges, S. Pailhès, Y. Sidis, A. Ivanov, C.~D. Frost, T.~G.
  Perring, C.~T. Lin, D.~P. Chen, and B. Keimer, Nature Physics {\bf 3},  780
  (2007).

\bibitem{vojta-04prl127002}
M. Vojta and T. Ulbricht, Phys. Rev. Lett {\bf 93},  127002  (2004).

\bibitem{uhrig-04prl267003}
G.~S. Uhrig, K.~P. Schmidt, and M. Gr\"{u}ninger, Phys. Rev. Lett {\bf 93},
  267003  (2004).

\bibitem{anisimov-04prb172501}
V.~I. Anisimov, M.~A. Korotin, A.~S. Mylnikova, A.~V. Kozhevnikov, D.~M.
  Korotin, and J. Lorenzana, Phys. Rev. B {\bf 70},  172501  (2004).

\bibitem{sachdev-90prb9323}
S. Sachdev and R.~N. Bhatt, Phys. Rev. B {\bf 41},  9323  (1990).

\bibitem{krueger-03prb134512}
F. Kr\"uger and S. Scheidl, Phys. Rev. B {\bf 67},  134512  (2003).

\bibitem{eder98prb12832}
R. Eder, Phys. Rev. B {\bf 57},  12832  (1998).

\bibitem{gopalan-94prb8901}
S. Gopalan, T.~M. Rice, and M. Sigrist, Phys. Rev. B {\bf 49},  8901  (1994).

\bibitem{tworzydlo-99prb115}
J. Tworzyd\l{}o, O.~Y. Osman, C.~N.~A. van Duin, and J. Zaanen, Phys. Rev. B
  {\bf 59},  115  (1999).

\bibitem{dalosto-00prb928}
S. Dalosto and J. Riera, Phys. Rev. B {\bf 62},  928  (2000).

\bibitem{nunner-02prb180404}
T.~S. Nunner, P. Brune, T. Kopp, M. Windt, and M. Gr\"uninger, Phys. Rev. B
  {\bf 66},  180404(R)  (2002).

\bibitem{seibold-05prl107006}
G. Seibold and J. Lorenzana, Phys. Rev. Lett. {\bf 94},  107006  (2005).

\bibitem{seibold-06prb144515}
G. Seibold and J. Lorenzana, Phys. Rev. B {\bf 73},  144515  (2006).

\bibitem{greiter-10prb144509}
M. Greiter and H. Schmidt, Phys. Rev. B {\bf 82},  144512  (2010).

\bibitem{yao-06prl017003}
D.~X. Yao, E.~W. Carlson, and D.~K. Campbell, Phys. Rev. Lett. {\bf 97},
  017003  (2006).

\bibitem{sachdev92prb12377}
S. Sachdev, Phys. Rev. B {\bf 45},  12377  (1992).

\bibitem{greiter02prb134443}
M. Greiter, Phys. Rev. B {\bf 65},  134443  (2002).

\bibitem{rice-97prb14655}
T.~M. Rice, S. Haas, M. Sigrist, and F.-C. Zhang, Phys. Rev. B {\bf 56},  14655
   (1997).

\bibitem{greiter10ap1349}
M. Greiter, Ann.\ Phys.\ {\bf 325},  1349  (2010).

\bibitem{anderson52pr694}
P.~W. Anderson, Phys. Rev. {\bf 86},  694  (1952).

\end{thebibliography}

\end{document}